\documentclass[%
 reprint,
 superscriptaddress,
 onecolumn,
 amsmath,amssymb,
 aps,
]{revtex4-2}

\pdfoutput=1

\usepackage{xcolor}					

\usepackage{placeins}
\usepackage{float}
\usepackage{graphicx}
\usepackage{gensymb}
\usepackage{amsmath,amssymb}
\usepackage{placeins}
\usepackage{multirow}
\usepackage{booktabs}
\usepackage[english]{babel}

\usepackage{subfigure}
\usepackage{overpic}
\usepackage{xfrac}

\begin{document}

\title{Large-scale motions and self-similar structures in compressible turbulent channel flows}

\author{Cheng Cheng}
 \affiliation{Department of Mechanical and Aerospace Engineering, The Hong Kong University of Science and Technology, Clear Water Bay, Kowloon, Hong Kong}

\author{Lin Fu}
\email{Corresponding author. Email: linfu@ust.hk}
 \affiliation{Department of Mechanical and Aerospace Engineering, The Hong Kong University of Science and Technology, Clear Water Bay, Kowloon, Hong Kong}
 \affiliation{Department of Mathematics, The Hong Kong University of Science and Technology, Clear Water Bay, Kowloon, Hong Kong}
\affiliation{HKUST Shenzhen-Hong Kong Collaborative Innovation Research Institute, Futian, Shenzhen, China}

\date{\today}

\begin{abstract}
In this work, we study the scale characteristics of the log- and outer-region motions and structures in subsonic and supersonic turbulence. 
To this end, a series of direct numerical simulations of the compressible turbulent channel flow at medium Reynolds numbers are performed. 
Based on this database, the streamwise and spanwise length scales
of the outer-region motions are investigated by the two-point correlations and the one-dimensional spectra. 
The energy distribution among the multi-scale structures in the outer region is found to be dominated by
the semilocal friction-Reynolds-number effects rather than the Mach-number effects.
This conclusion not only holds for the velocity fluctuations but also the fluctuations of the thermodynamic variables. 
Besides, the streamwise and spanwise length scales of the outer motions do not alter significantly when the flow passes the sound barrier as reported by a previous experimental study (Bross et al., J. Fluid
Mech., vol. 911, 2021, A2). 
On the other hand, the self-similar structures populating the logarithmic region are investigated by adopting a linear coherence spectrum. 
The streamwise/wall-normal aspect ratio of the self-similar wall-attached structures of the streamwise velocity and temperature fluctuations is approximately 15.5, and the counterpart of density and pressure
fluctuations is 1.8. 
The present study confirms the existence of self-similar structures in compressible wall turbulence and assesses their geometrical characteristics.

\end{abstract}

\maketitle

\section{Introduction}
Since the discovery of the streaks in the near-wall region of the turbulent boundary layers by \citet{Kline1967}, the coherent structures, as well as the energy-containing motions (eddies) in wall-bounded turbulence have been extensively investigated over the past decades. 
Now, it is understood that in addition to the streaks and the vortical structures residing in the near-wall region \citep{Kline1967,Schoppa2002}, there are also large-scale motions (LSMs), very-large-scale motions (VLSMs), and self-similar motions populating the logarithmic and outer regions. 
These three typical energy-containing motions have been unmasked in high-Reynolds-number pipes \citep{Kim1999,Wu2012}, channels \citep{DelAlamo2003,Lozano-Duran2014a}, and zero-pressure-gradient turbulent boundary layers \citep{Sillero2013,Sillero2014}.
Besides, they have also been identified in compressible wall-bounded turbulence \citep{Ganapathisubramani2006,Modesti2016,Bross2021}. 
Further studies have also revealed that these energy-containing motions not only exert pronounced influences on the near-wall turbulent intensities \citep{Mathis2009,Baars2016}, but also the unsteadiness of the separated bubble downstream of the shockwave/boundary-layer interactions \citep{Ganapathisubramani2009,Baidya2020}. 
Thus, the investigation of their physical characteristics has remarkable significance for drag reduction and flow-separation control of the high-speed vehicles. 
It should be noted that in internal flows, the long streaky low-streamwise-momentum 
motions are often referred to as `VLSMs', whereas in external flows they are more commonly
referred to as `superstructures' \citep{Smits2011}. In the present study, we do not differentiate between the two terminologies and denote these motions in internal/external flows as VLSMs collectively.

LSMs have a streamwise scale of approximately $1-2\delta$ ($\delta$ indicates the boundary thickness), and VLSMs are generally believed to extend several $\delta_{}$ long along the streamwise direction (the exact length is disputed) and meander in spanwise direction \citep{Smits2011}.
The investigations of the incompressible wall-bounded turbulence have provided the majority of the knowledge concerning these motions. It is speculated that the coherent-structure organizations of compressible wall turbulence in the logarithmic and outer regions would resemble those of incompressible flows. 
However, this hypothesis is still controversial. 

As early as 1989, \citet{Smits1989} compared the turbulent structures of subsonic ($Ma=0.1$, here $Ma$ denotes the Mach number) and supersonic ($Ma=2.9$) boundary layers, and found that their spanwise scales are almost identical, but the streamwise scales in the supersonic flow are about half the size of those in subsonic flow. Their analyses are based on the space-time correlations of the mass-flux fluctuations measured from the hot-wire (HW) probes. \citet{Smits2006} also summarized the experimental measurements (hot-wire measurements dominate) and came to the conclusion that the streamwise length scales of outer motions decrease obviously  with increasing Mach number, whereas their spanwise scales have no statistical changes with the Reynolds and Mach numbers. \textcolor{black}{However, \citet{Spina1994} reported that the average spanwise extent of the largest eddies in a Mach 3 turbulent boundary layer is similar to that of subsonic turbulent boundary layers by adopting hot-wire and flow visualizations, whereas their average streamwise scales are about twice those of low Reynolds number, subsonic turbulent boundary layers.}

Nevertheless, these propositions are not totally in accordance with the measurement via particle image velocimetry (PIV). \citet{Ganapathisubramani2006} used planar PIV to investigate the scale characteristics of the energy-containing motions in a Mach 2 turbulent boundary layer. By employing the two-point correlation, they discovered that the streamwise length scale of the streamwise velocity fluctuations at the outer region is nearly four-time long as that in incompressible cases, and their spanwise length scale is slightly larger than that of the subsonic flow. \textcolor{black}{However, \citet{Williams2018} also investigated a hypersonic turbulent boundary layer at Mach 7.5 by PIV, and their conclusion is different, e.g., the streamwise correlation lengths are less sensitive to the compressibility.}
Very recently,  \citet{Bross2021} examined a sequence of compressible turbulent boundary layers varying from subsonic to supersonic, by using planar 2-D PIV and stereo-PIV, and they observed that
there is a scale expansion in both streamwise and spanwise directions when flow passes the sound barrier. 
It's interesting to note that, rather than the absolute value of the Mach number, whether this phenomenon takes place relies on whether the flow speed is sub- or supersonic.

Direct numerical simulation (DNS) is another tool to tackle this question.
However, the findings from DNS diverge significantly from those of the aforementioned experimental studies.  \citet{Ringuette2008}  performed a DNS study of a Mach 3 turbulent boundary layer and found that VLSMs in the compressible flow are similar to those in incompressible cases, and their length scale is mostly shaped by the Reynolds-number effects. 
\citet{Pirozzoli2011} analyzed the structural organization of a Mach 2 turbulent boundary layer and reported that their streamwise length scales are identical to the incompressible cases, whereas the spanwise length scales are slightly larger. 
Later, they also observed that the spanwise length scales of the motions in the outer region of the Mach 1.5 turbulent channel flows are close to the incompressible flows \citep{Modesti2016}. \textcolor{black}{The scale invariance of LSMs in  supersonic wall turbulence has been drawn in a series of works of their group and collaborators \citep{Pirozzoli2011,Pirozzoli2012,Modesti2016,Cogo2022}.}
\textcolor{black}{Recently, similar results are also provided by \citet{Yao2020} and \citet{Huang2022} after examining their own DNS database of compressible channel flows and supersonic/hypersonic turbulent boundary layers with different thermal boundary conditions. The latter study also reveals that the wall cooling condition plays a minor role in shaping the length scales of LSMs and VLSMs.} \textcolor{black}{Table \ref{tab:refpaper} summarizes  the experimental and numerical studies on the length scales of the outer-region motions in compressible wall-bounded turbulence. Reviewing the works of predecessors, it can be found that the scale characteristics of the motions in the outer region of the compressible wall turbulence are ambiguous from the experimental and numerical sides. Most of the DNS studies hold the view that their length scales are not sensitive to compressibility, whereas for the experimental studies, it has been a question of different opinions.}

\begin{table}[!htbp]
\centering
\begin{tabular}{cccccccc}
\hline
\hline
Reference  & Year & Flow&  Ma & Method & $\lambda_x$ & $\lambda_z$ & Note \\ 
\hline
\hline
\citet{Smits1989}& 1989 & TBL& 0.1-2.9 & HW & $\searrow$ & $\rightarrow$ & None \\
\hline
\citet{Spina1994}& 1994 & TBL& 3 & HW & $\nearrow$ & $\rightarrow$ & None \\
\hline
\citet{Smits2006}&  2006& TBL& 0.035-11 & HW & $\searrow$ & $\rightarrow$ &  HW dominant\\
\hline
\citet{Ganapathisubramani2006}& 2006& TBL & 2.0 & PIV & $\nearrow$ & $\nearrow$ & None \\
\hline
\citet{Williams2018}& 2018& TBL & 7.5 & PIV & $\rightarrow$ & None & None \\
\hline
\citet{Bross2021}& 2021 & TBL& 0.3-3 & PIV & $\nearrow$ & $\nearrow$ & Transonic expansion \\
\hline
\citet{Pirozzoli2011} & 2011& TBL  & 2 & DNS  &  $\rightarrow$ & $\nearrow$ & None \\
\hline
\citet{Pirozzoli2012} & 2012 & TBL/CP  & 0.3-4 & DNS  &  None & $\rightarrow$ & None \\
\hline
\citet{Modesti2016} & 2016& CH  & 0.1-3 & DNS  &  None & $\rightarrow$ & None \\
\hline
\citet{Yao2020} & 2020& CH  & 0-1.5 & DNS  &  $\rightarrow$ & $\rightarrow$ & None \\
\hline
\citet{Huang2022} & 2022& TBL  & 2.5-10.9 & DNS  &  $\rightarrow$ & $\rightarrow$ & None \\
\hline
\citet{Cogo2022} & 2022& TBL  & 2-5.86 & DNS  &  None & $\rightarrow$ & None \\
\hline
\end{tabular}
\caption{\textcolor{black}{Summary of the experimental and numerical studies on the length scales of the outer-region motions in compressible wall-bounded turbulence. Here, TBL, CP, and CH denote turbulent boundary layer, Couette-Poiseuille flow, and channel flow, respectively. $\lambda_x$ and $\lambda_z$ denote the streamwise and spanwise length scales, respectively. The symbols $\nearrow$ and $\searrow$ denote larger and smaller than the incompressible/subsonic counterparts, respectively, whereas $\rightarrow$ denotes no change.}
\label{tab:refpaper}}
\end{table}

On the other hand, it is documented  that there are self-similar motions populating the logarithmic region of the wall turbulence \citep{Baars2017,Cheng2020,Cheng2022,Cheng2022a}. 
According to the well-known attached-eddy model, the logarithmic region is occupied by an array of self-similar energy-containing motions (or eddies) with their roots attached to the near-wall region \citep{Townsend1976,Perry1982}. 
Hence, examining the geometrical features of these self-similar structures is beneficial for both developing the attached-eddy hypothesis and advancing the knowledge of turbulent dynamics. 
In this respect, nearly all the works concentrate on incompressible wall-bounded turbulence. 
For example, \citet{Baars2017} identified the self-similar structures
of the streamwise velocity fluctuations in incompressible boundary layers by a linear coherence spectrum and found the streamwise/wall-normal aspect ratio ($AR$)
of them is roughly 14. 
This scale characteristic is then employed to construct a filter for isolating the spectral energy results from the wall-attached eddies \citep{Baars2020a}. 
Whether it still holds in compressible turbulence? Whether the thermodynamic variables bear the self-similar characteristics as the streamwise velocity fluctuations? If yes, what is their $AR$? None of these issues have been reported to be thoroughly investigated. \textcolor{black}{Very recently, \citet{Yu2022a} employed the proper orthogonal decomposition to identify the self-similar structures of the streamwise velocity and the temperature fluctuations in a compressible channel flow, but their geometrical characteristics have not been clarified.}

The work of the present study is divided into two parts. In the first part, we conduct a series of DNSs of the subsonic and the supersonic turbulent channel flows at medium Reynolds numbers (the maximal friction Reynolds number is 1150). Then, in conjunction with the open-source incompressible data, the Mach-number effects on the scale characteristics of the outer-region motions are dissected, together with the scale characteristics of the thermodynamic variables, such as $p'$ (pressure fluctuation), $T'$ (temperature fluctuation), and $\rho'$ (density fluctuation). Particular attention is paid to the scale expansion when flow passes the sound barrier reported by a recent study \citep{Bross2021}. In the second part, we elaborate on the geometrical characteristics of the self-similar structures populating the logarithmic region. We generalize the work of \citet{Baars2017} to the compressible flows by leveraging the database set up in the present study. We not only compare the geometrical properties of the self-similar wall-attached velocity fluctuation with the incompressible counterparts but also unravel those of the thermodynamic variables. The present work may help develop an advanced scale-resolving method for predicting the compressible wall-bounded flows \cite{fu2021shock, fu2022prediction}.

\textcolor{black}{The focuses of the present study are summarized as follows.}

\textcolor{black}{(1) The scale characteristics of the outer-region motions in subsonic and supersonic wall turbulence, not only for the velocity fluctuations, but also the fluctuations of the thermodynamic variables. The latter has rarely been reported before. Particular attention is paid to the scale expansion when flow passes the sound barrier reported by a recent study \citep{Bross2021}.}

\textcolor{black}{(2) The possible factors that may account for the differences between the experimental and numerical studies, such as the 
variable for study, the method to normalize the spectrum, and the wall-normal position of the detected plane.}

\textcolor{black}{(3) The scale characteristics of the self-similar structures in the logarithmic region, not only for the streamwise velocity fluctuations, but also the thermodynamic variables.}

The paper is organized as follows: In Sec. \ref{DNS}, the DNS setup and numerical validation are provided in detail. The  
scale characteristics of the outer-region motions in compressible channel flow are analyzed by the 
two-point correlation and the one-dimensional spectrum in Sec. \ref{RD}, concurrently, the geometrical characteristics of the self-similar structures in the logarithmic region are also investigated in this section. Conclusions are given in Sec. \ref{CO}.  

\section{NUMERICAL setup and validation}\label{DNS} 
DNSs of compressible turbulent channel flows
have been conducted with a finite-difference code, by solving the three-dimensional
unsteady compressible Navier–Stokes equations. 
The convective terms are discretized
with a seventh-order upwind-biased scheme, and the viscous terms are evaluated with an eighth-order central difference scheme. 
Time advancement is performed using the third-order strong-stability-preserving (SSP) Runge–Kutta method \citep{Gottlieb2001}. 
A constant molecular Prandtl
number $Pr$ of 0.72 and a specific heat ratio $\gamma$ of 1.4 are employed. The dependence of dynamical viscosity $\mu$ on temperature $T$ is given by Sutherland's law, i.e.,
\begin{equation}
\mu=\mu_{0} \frac{T_{0}+S}{T+S}\left(\frac{T}{T_{0}}\right)^{3 / 2},
\end{equation}
where $S=110.4K$ and $T_0=273.1K$.

All the DNSs are carried out in a rectangular box with its sizes along streamwise ($x$), spanwise ($z$), and wall-normal ($y$) directions denoted as $L_x$, $L_z$, and $L_y$, respectively. 
In the streamwise and spanwise directions, the mesh is uniformly spaced, whereas, in the wall-normal direction, the mesh is hyperbolically clustered towards the walls. 
The isothermal
no-slip conditions are imposed at the top and bottom walls, and the periodic boundary condition is imposed in the wall-parallel directions, i.e., $x$ and $z$ directions. 
All simulations begin with a parabolic velocity profile with random perturbations superimposed and uniform temperature and density values. A body force is imposed in the streamwise direction to maintain a constant
mass flow rate, and a corresponding source term is also added in the energy equation \citep{Huang1995}.

In the present study, we carry out two simulations at a bulk Mach number $M_b=U_b/C_w=0.8$ ($U_b$ is the bulk velocity, and $C_w$ is the speed of sound at wall temperature) and $Re_b=\rho_bU_bh/\mu_w=7667$ and $17000$ ($\rho_b$ denotes the bulk density, $h$ the channel half-height, and $\mu_w$ the dynamic
viscosity at the wall). 
Two other DNSs at a bulk Mach number $M_b=1.5$ and $Re_b=9400$ and $20020$ are also conducted. For all cases, the computational domain has the same dimensions, namely, $L_x\times L_z\times L_y=4\pi h\times 2\pi h \times 2h$.
Previous studies have verified that these setups of dimensions can capture most of the large-scale motions in the outer region of the boundary layer \citep{Agostini2014,Agostini2019}. 
 Details of the parameter settings are listed in Table \ref{tab:grid}. 
Two incompressible cases at $Re_{\tau}=547$ and $Re_{\tau}=934$ by \citet{DelAlamo2003} and \citet{DelAlamo2004} are also employed for comparison. Details of the parameter settings are listed in Table \ref{tab:grid2}.
\begin{table}[!htbp]
\centering
\begin{tabular}{cccccccccc}
\hline
\hline
Case  & $M_b$ & $Re_b$ & $Re_{\tau}$ & $Re_{\tau}^*$ &$\Delta x^+$ & $\Delta z^+$ & $\Delta y_{min}^+$ & $\Delta y_{max}^+$ &  $Tu_{\tau}/h$ \\ 
\hline
\hline
Ma08Re8K  & 0.8 & 7667 & 436 & 382 &  10.8 &   6.9 & 0.44 &  5.4 &  49.4 \\
\hline
Ma08Re17K & 0.8 & 17000 & 882 & 778 &  10.8 &  6.5 & 0.63 & 6.4  &  15.3  \\
\hline
Ma15Re9K & 1.5 & 9400 & 594 & 395 & 7.3 &   3.7 &  0.5 &   5.9   &  30.2  \\
\hline
Ma15Re20K & 1.5 & 20020 & 1150 & 780 &  9.3 & 4.7 &  0.49& 6.9   &  9.1  \\
\hline
\end{tabular}
\caption{Parameter settings of the compressible DNS database. Here, $M_b$ denotes the bulk Mach number.
    $Re_b$, $Re_{\tau}$ and $Re_{\tau}^*$ denote the bulk Reynolds number, friction Reynolds number, and
    semilocal friction Reynolds number, respectively.
        $\Delta x^+$ and $\Delta
		z^+$ denote the streamwise and spanwise grid
		resolutions in viscous units, respectively. $\Delta
		y_{min}^+$ and $\Delta y_{max}^+$ denote the finest and
		the coarsest resolution in the wall-normal direction,
		respectively. $Tu_{\tau}/h$ indicates the total eddy turnover time used to accumulate statistics.\label{tab:grid}}
\end{table}
\begin{table}[!htbp]
\centering
\begin{tabular}{ccccccccccc}
\hline
\hline
Case   & $Re_{\tau}$ & $Re_{\tau}^*$ &   $L_x(h)$&   $L_y(h)$ & $L_z(h)$  &$\Delta x^+$ & $\Delta z^+$ & $\Delta y_{min}^+$ & $\Delta y_{max}^+$ &  $Tu_{\tau}/h$ \\ 
\hline
\hline
Ma00Re10K  & 547& 547 & $8\pi$ & 2& $4\pi$ & 13.4 & 6.8 & 0.04 & 6.7  & 22\\
\hline
Ma00Re18K & 934& 934 & $8\pi$ &2& $3\pi$  & 11.5 & 5.7 & 0.03 & 7.6  & 12\\
\hline
\end{tabular}
\caption{Parameter settings of the incompressible DNS database.\label{tab:grid2}}
\end{table}

Both the Reynolds (denoted as $\bar{\phi}$) and the Favre averaged (denoted as $\widetilde{\phi}=\overline{\rho\phi}/\overline{\rho}$) statistics are used in the present study. 
The corresponding fluctuating components are represented as $\phi'$ and $\phi''$, respectively.
 Hereafter, we use the superscript $+$ to represent the normalization with the friction velocity (denoted as $u_{\tau}$, $u_{\tau}=\sqrt{\tau_w/\rho_w}$, $\tau_w$ is the mean wall-shear stress), and the viscous length scale (denoted as $\delta_{\nu}$, $\delta_{\nu}=\nu_w/u_{\tau}$, $\nu_w=\mu_w/\rho_w$). 
 We also use the superscript $*$ to represent the normalization with the semilocal wall units, i.e., $u_{\tau}^*=\sqrt{\tau_w/\overline{\rho}}$ and $\delta_{\nu}^*=\overline{\nu(y)}/u_{\tau}^*$. 
 Hence, the relationship between the semilocal friction Reynolds number and the friction Reynolds number is $R e_{\tau}^{*}=R e_{\tau} \sqrt{\left(\overline{\rho_c} / \bar{\rho}_{w}\right)} /\left(\overline{\mu_c} / \bar{\mu}_{w}\right)$. The subscript $c$ refers to the quantities evaluated at the channel center.
 It is noted that the cases Ma08Re8K and Ma08Re17K share similar $Re_{\tau}^*$  with the cases Ma15Re9K and Ma15Re20K, respectively.

\begin{figure*}
    \centering
    \includegraphics[width=0.45\linewidth]{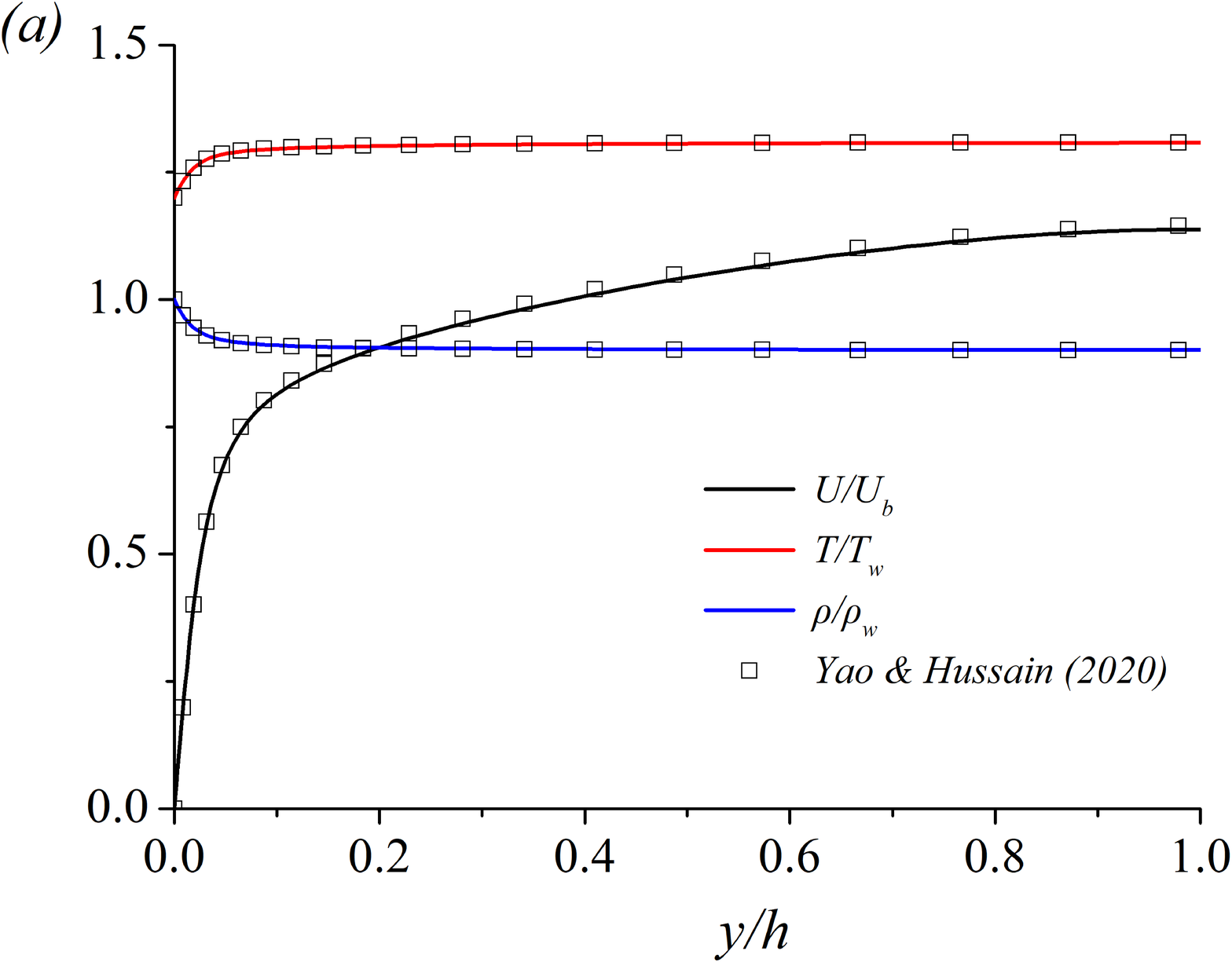}
    \includegraphics[width=0.45\linewidth]{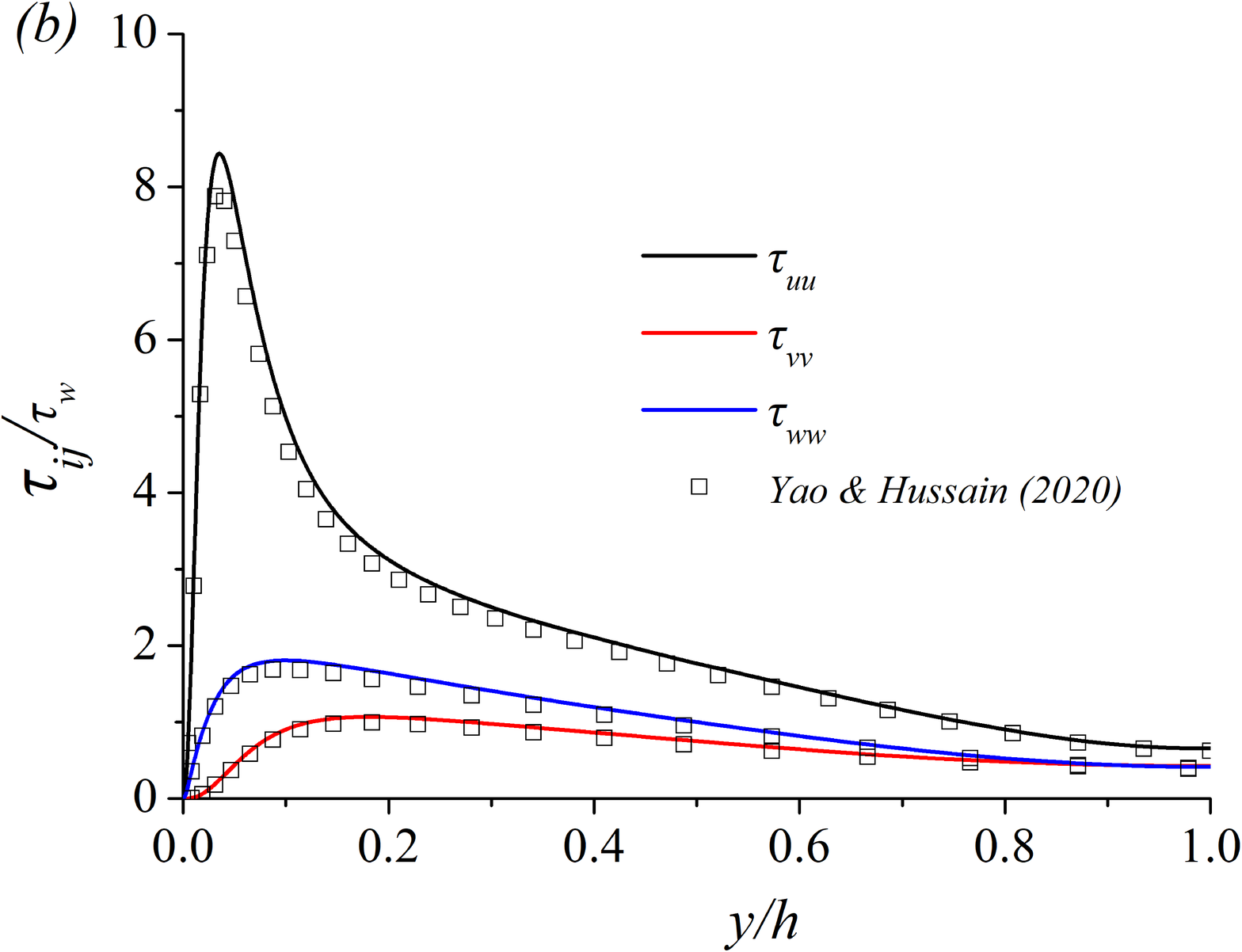} 
    \includegraphics[width=0.45\linewidth]{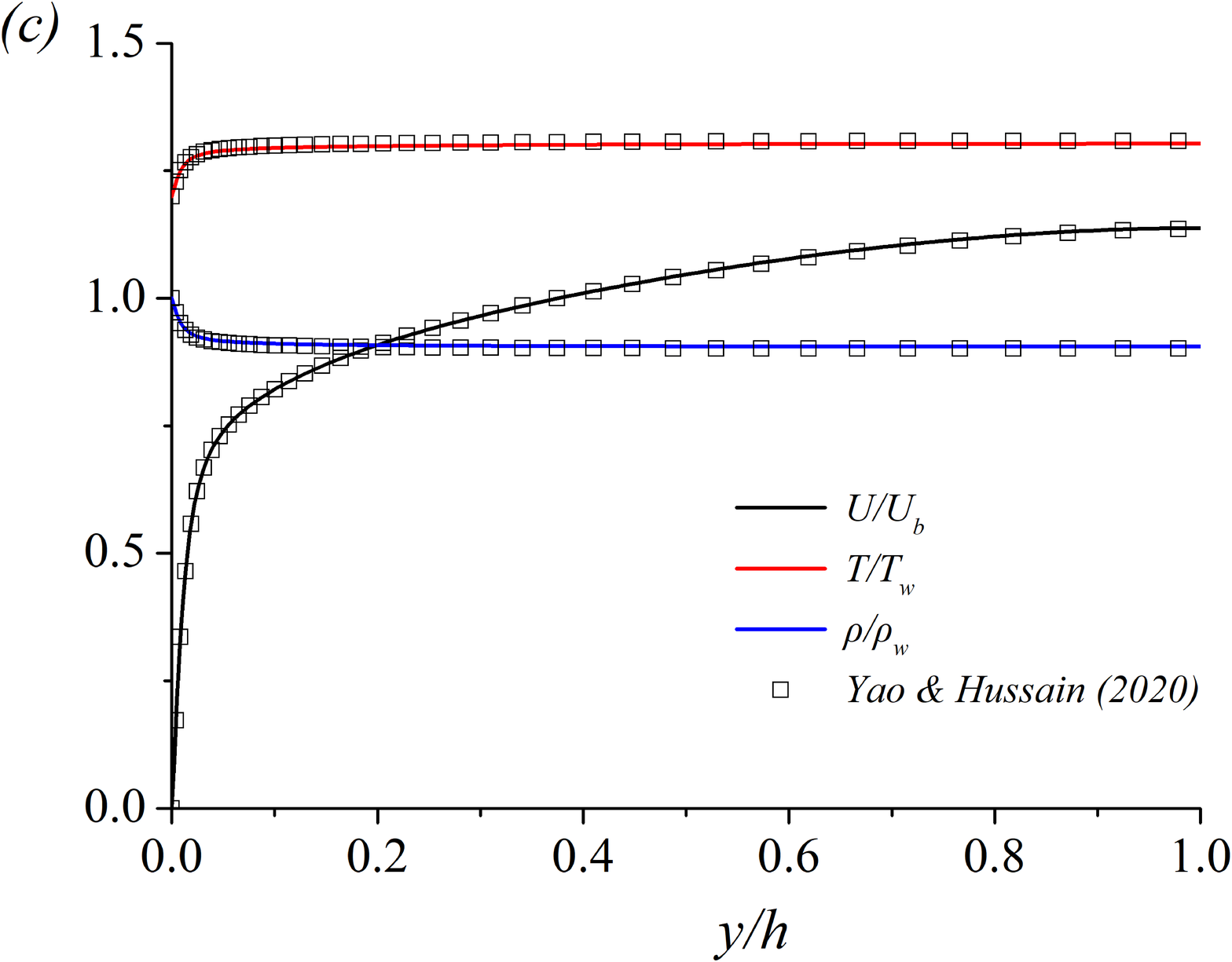}
    \includegraphics[width=0.45\linewidth]{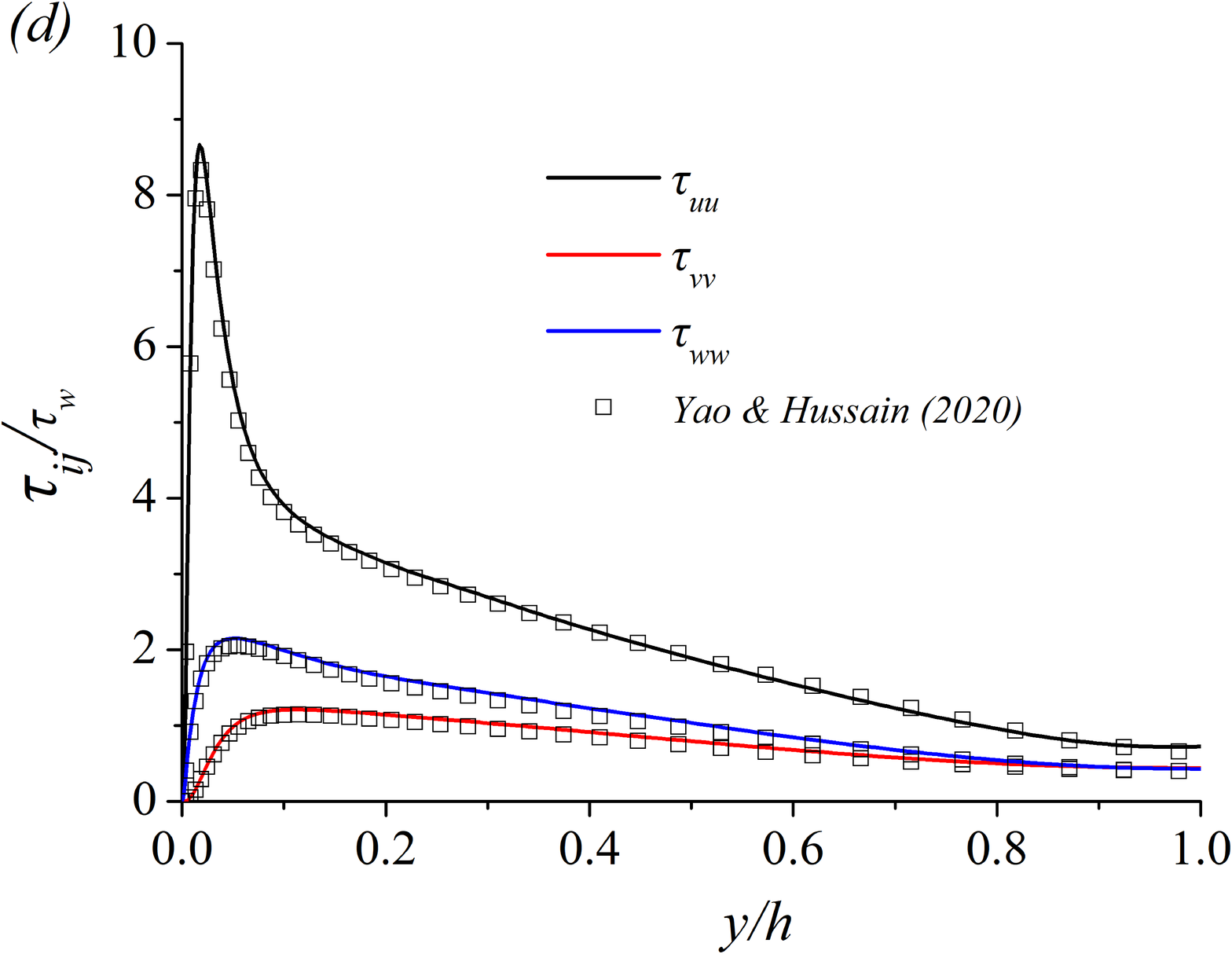} 
    \caption{Profiles of mean streamwise velocity, mean temperature, and mean density for the
    cases Ma08Re8K ($a$) and Ma08Re17K ($c$); profiles of the Reynolds stress for the cases Ma08Re8K ($b$) and Ma08Re17K ($d$). The profiles of $T/T_w$ are shifted upward by $\Delta (T/T_w)=0.2$ for better comparison. }
    \label{fig:valid1}
\end{figure*}
%
\begin{figure*}
    \centering
    \includegraphics[width=0.45\linewidth]{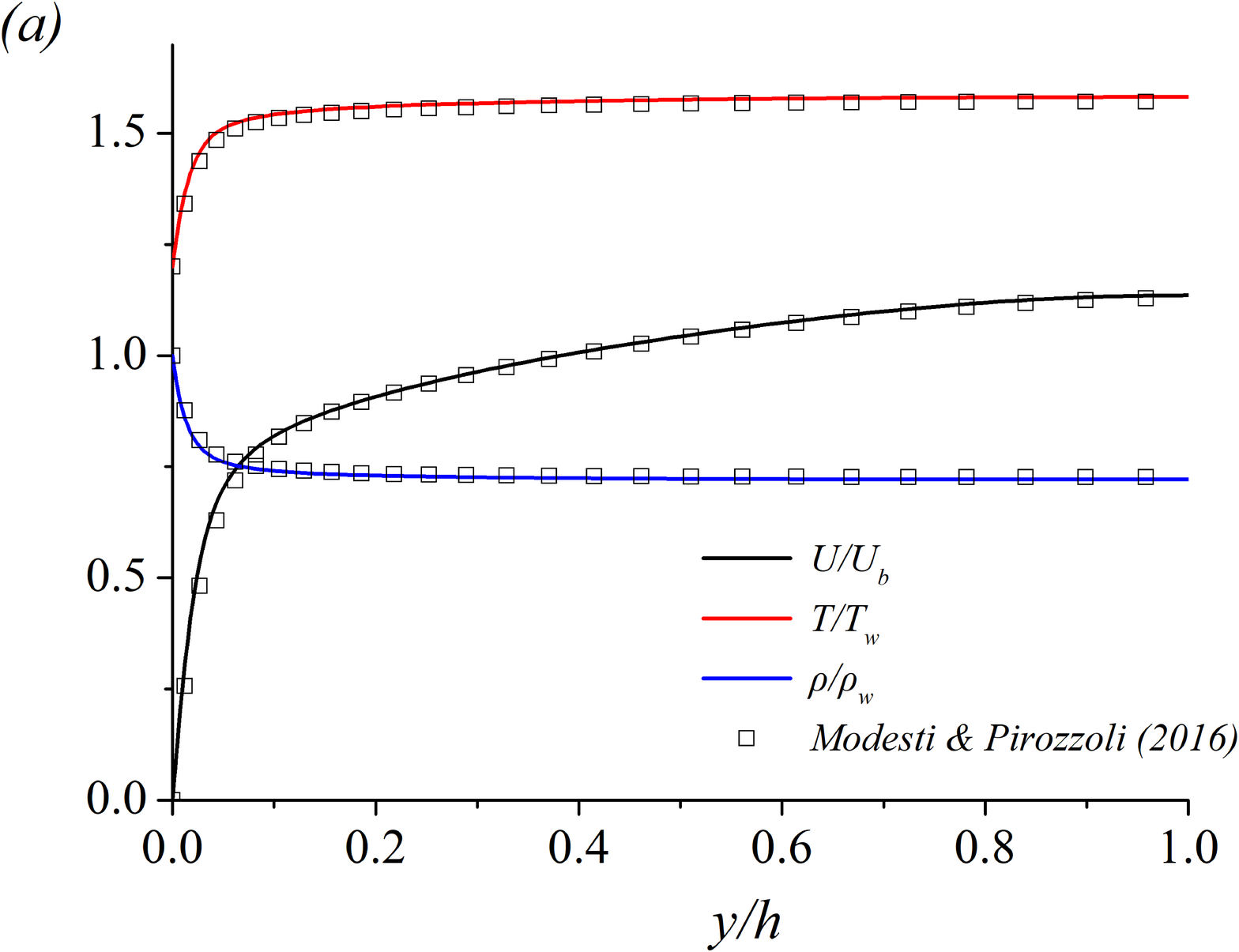}
    \includegraphics[width=0.45\linewidth]{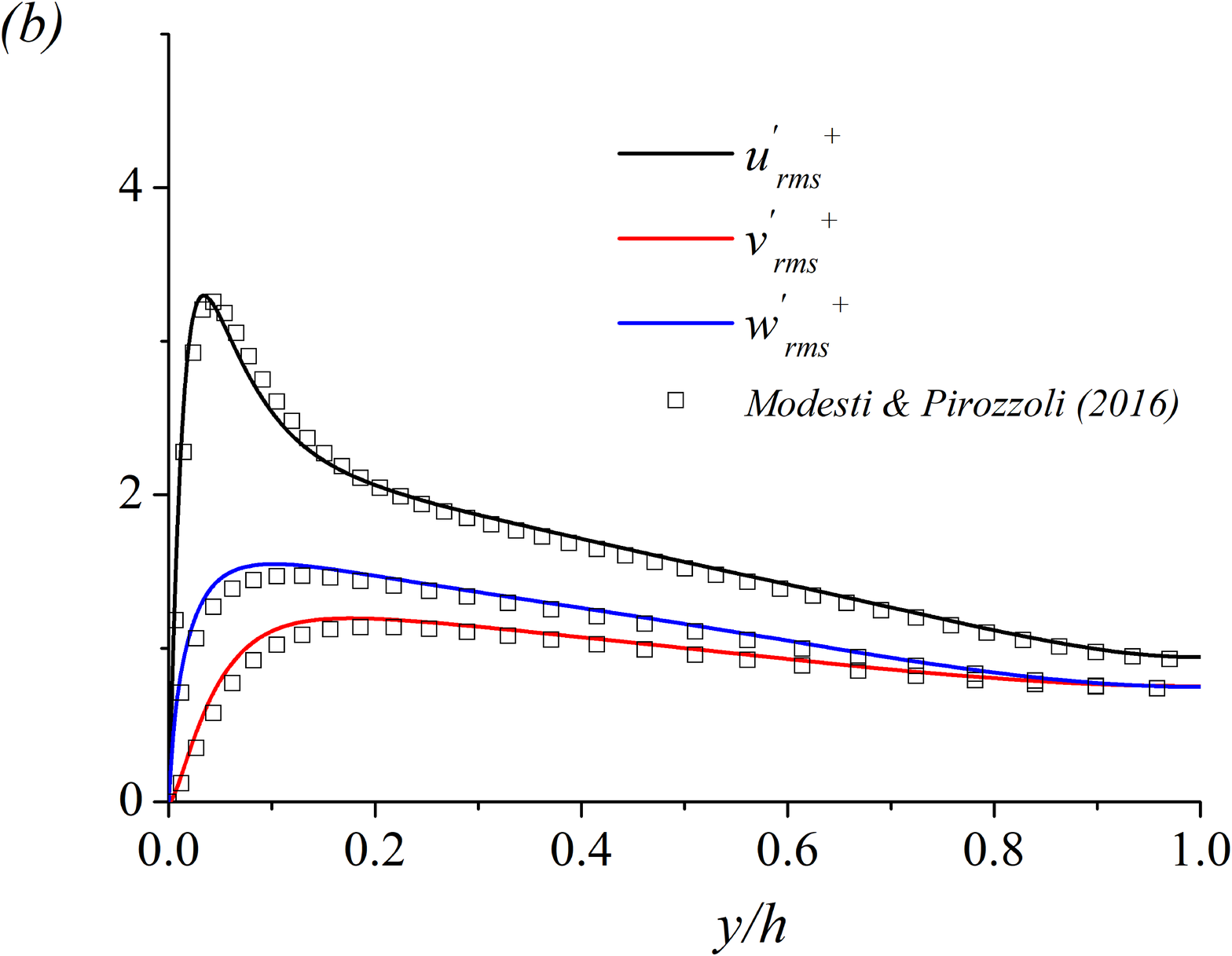} 
    \includegraphics[width=0.45\linewidth]{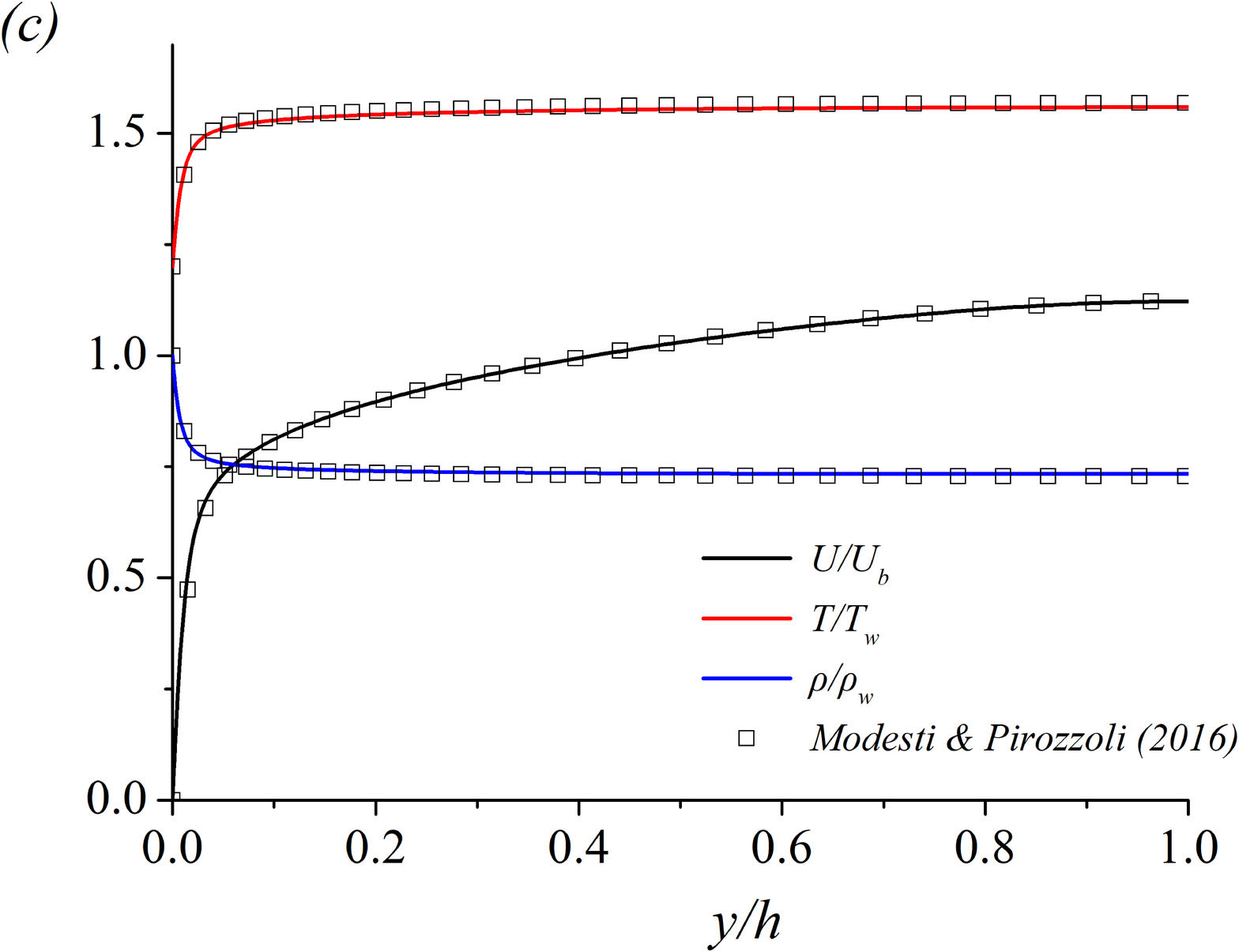}
    \includegraphics[width=0.45\linewidth]{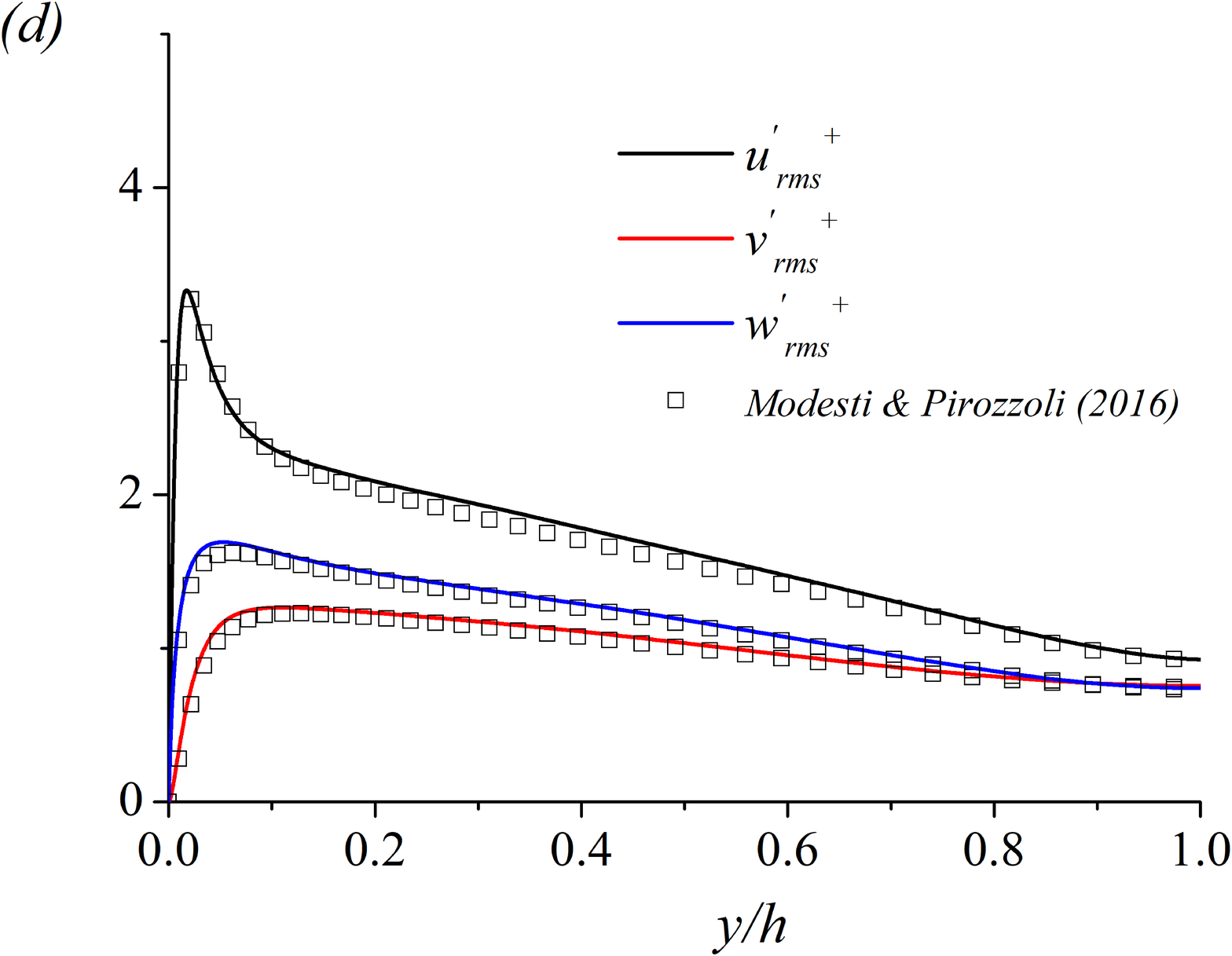} 
    \caption{Profiles of mean streamwise velocity, mean temperature, and mean density for the
        cases Ma15Re9K ($a$) and Ma15Re21K ($c$); profiles of the root-mean-square of the velocity fluctuation for the cases Ma15Re9K ($b$) and Ma15Re21K ($d$). The profiles of $T/T_w$ are shifted upward by $\Delta (T/T_w)=0.2$ for better comparison.}
    \label{fig:valid2}
\end{figure*}
%
To verify the compressible dataset of present study, Fig.~\ref{fig:valid1} compares the DNS results of Ma08Re8K and Ma08Re17K with the flow statistics of \citet{Yao2020} at identical $Ma_b$ and $Re_{b}$, respectively. Both the mean quantities and the Reynolds stress $\tau_{i j}=\bar{\rho} R_{i j}$ with $R_{i j}=\widetilde{u_{i}^{\prime \prime} u_{j}^{\prime \prime}}=\widetilde{u_{i} u_{j}}-\widetilde{u_{i}} \widetilde{u_{j}}$ ($v$ and $w$ denote the wall-normal and spanwise velocity, respectively) are compared. Fig.~\ref{fig:valid2} compares the mean quantities and the root-mean-square (r.m.s.) of the velocity fluctuations of Ma15Re9K and Ma15Re20K with the results reported by \citet{Modesti2016} at $M_b=1.5$, $Re_b=7667$ and $M_b=1.5$, $Re_b=17000$, respectively. All the profiles of the concerned quantities agree reasonably with the previous studies. The minor differences shown in Fig.~\ref{fig:valid2} are ascribed to the distinctions in $Re_b$. All these confirm the accuracy of the present database. 
\begin{figure*}
    \centering
    \includegraphics[width=0.45\linewidth]{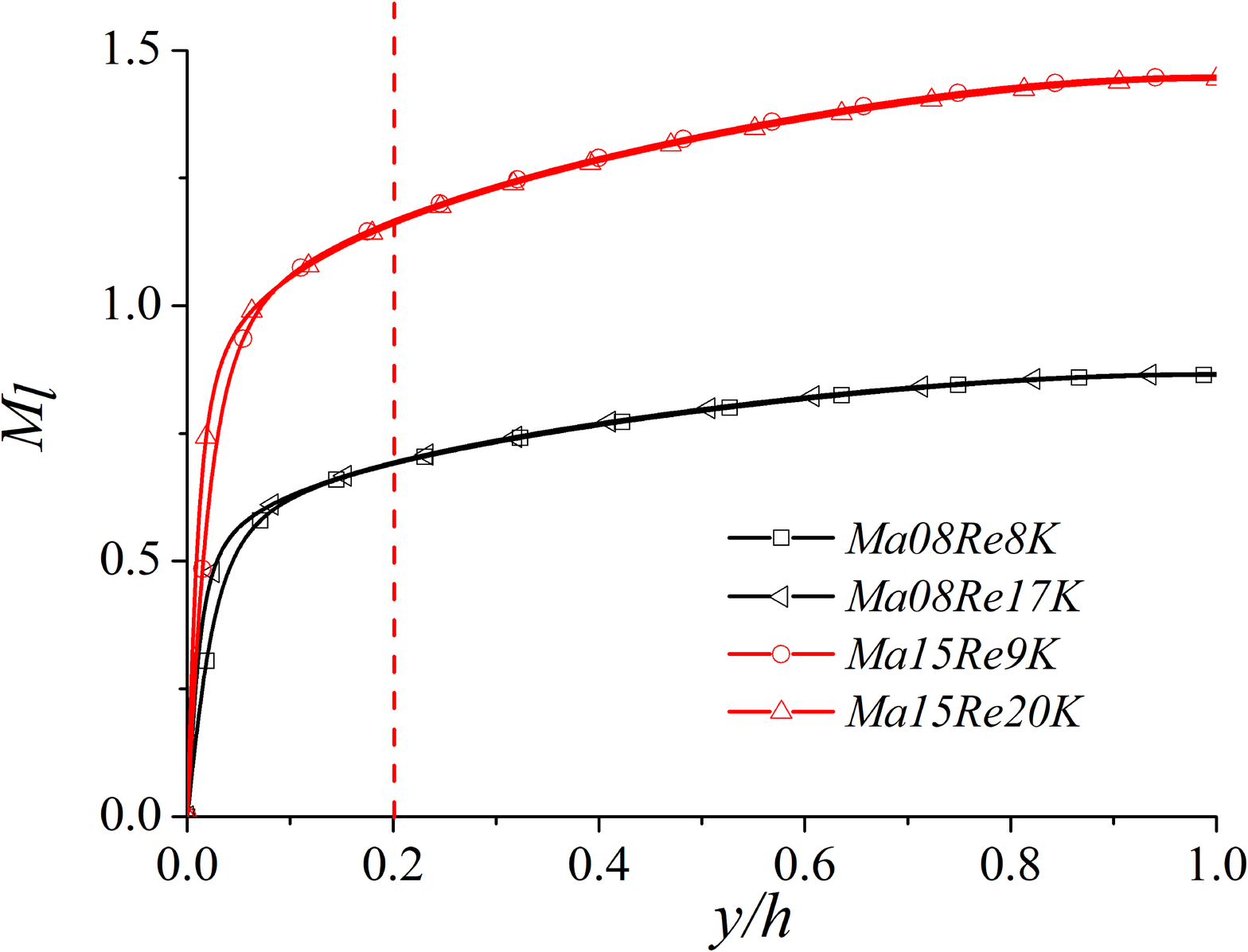}
    \caption{Profiles of mean local Mach number for all compressible cases. The red dashed line indicates $y=0.2h$, i.e., the wall-normal position for investigation in the present study. }
    \label{fig:Mal}
\end{figure*}
%

Fig.~\ref{fig:Mal} plots the variations of the mean local Mach number ($M_l$) as functions of the wall-normal height $y/h$. 
For the cases at $M_b=0.8$, $M_l\approx0.69$ at $y=0.2h$, whereas for the cases at $M_b=1.5$, $M_l\approx1.16$ at the same wall-normal height. 
Considering that $y=0.2h$ is the upper bound of the logarithmic region \citep{Jimenez2018}, dissecting the scale characteristics at this wall-normal height can shed light on the differences between the large-scale motions in the subsonic and supersonic wall-bounded turbulence. 
The results presented below are mainly focused on this wall-normal position. We also show that the conclusions drawn below are not changed with varying wall-normal heights within the outer region (see appendix A).

\section{Results and discussions} \label{RD}
\subsection{Variable for characterizing outer-layer motions}

Before processing our analysis, it is crucial to clarify which physical variable is appropriate to characterize the scales of the outer-layer motions. 
Several variables have been employed by previous studies to scrutinize the scale characteristics of compressible wall-bounded turbulence. 
For example, \citet{Smits1989} adopted the correlated signals of the mass-flux fluctuations $(\rho u)'$ to depict the spatial organizations of the coherent structures in the compressible boundary layers. \citet{Ganapathisubramani2006} used the two-point correlation of $u'$ to measure the length scale of VLSMs in a turbulent boundary layer of Mach 2. 
Moreover, $\sqrt\rho u^{''}$ has also been visualized to display the formation of the energy-containing motions vividly \citep{Yao2020,Hirai2021,Huang2022}.

\begin{figure*}
    \centering
    \includegraphics[width=0.45\linewidth]{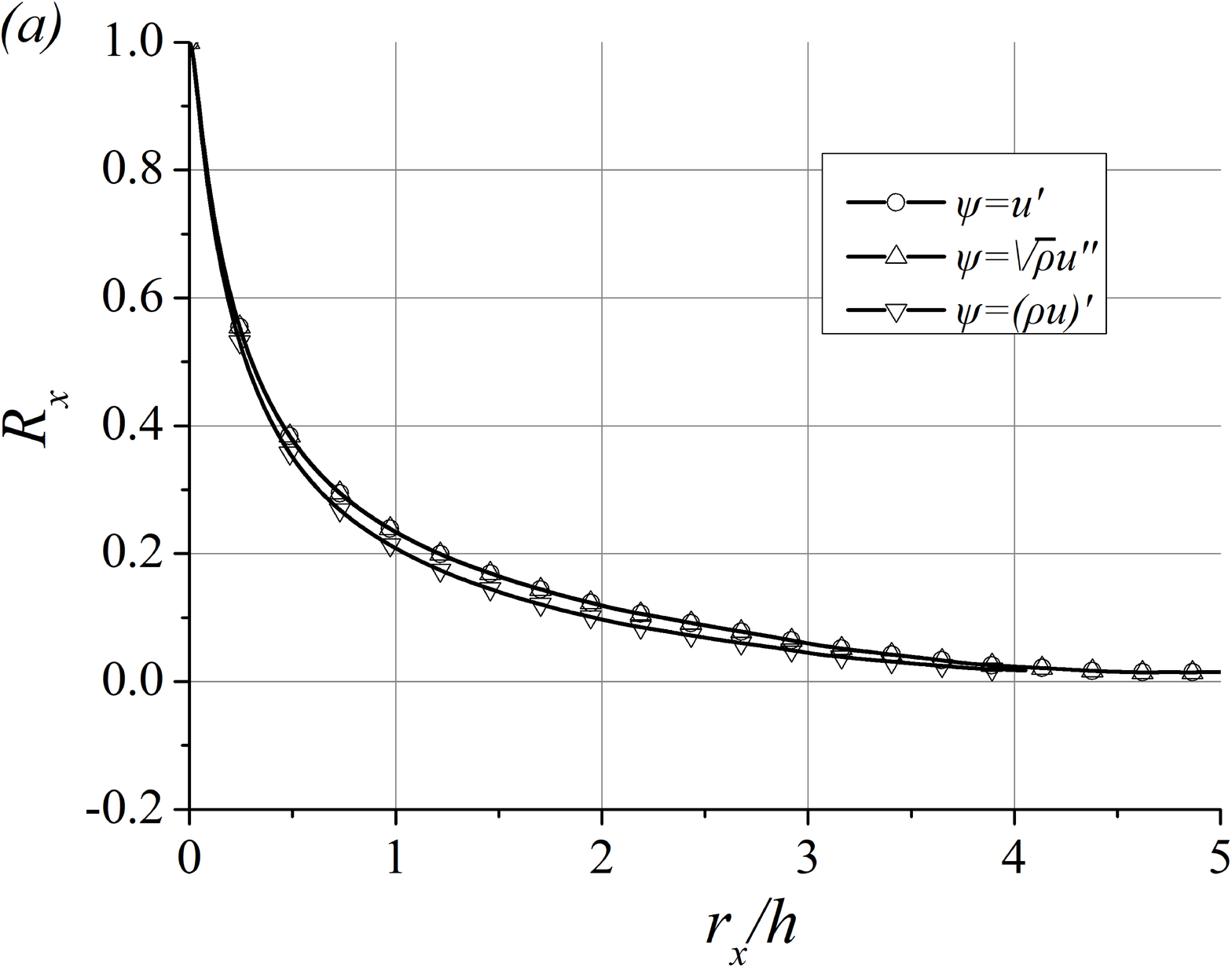}
    \includegraphics[width=0.45\linewidth]{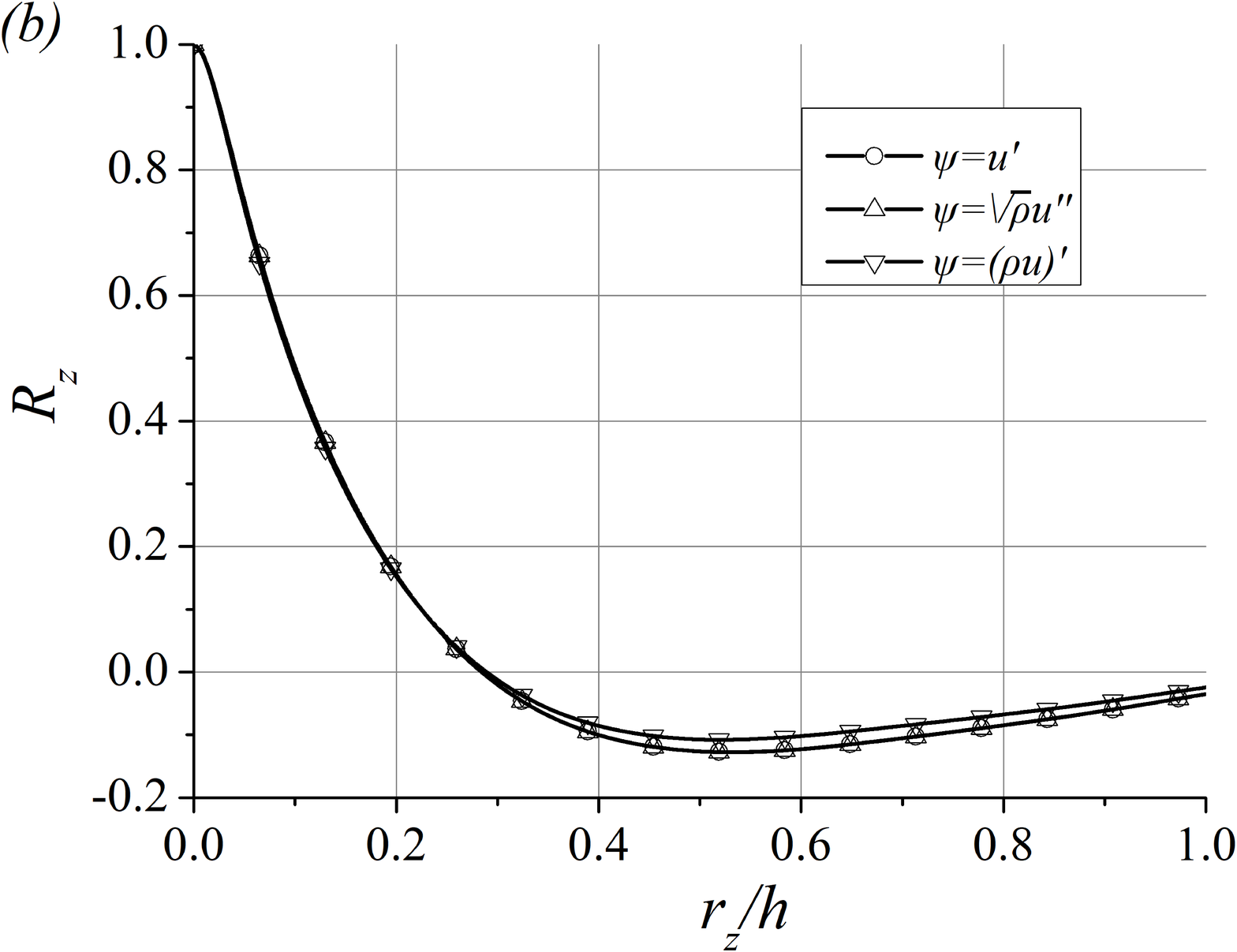}
    \caption{Streamwise ($a$) and  spanwise ($b$) two-point correlations of $u'$, $\sqrt\rho u^{''}$, and $(\rho u)'$ for the case Ma15Re20K at $y/h=0.2$.}
    \label{fig:DQ}
\end{figure*}

In this study, the behaviors of these quantities are compared by utilizing the streamwise and spanwise two-point correlations based on the case Ma15Re20K. For the streamwise two-point correlation of variable $\psi$, it takes the form of
\begin{equation}\label{eq:Rx}
R_{x}\left(r_{x}, y\right)=\frac{\left\langle \psi(x, y, z) \psi\left(x+r_{x}, y, z\right)\right\rangle }{\left\langle \psi^2\right\rangle},
\end{equation}
where $r_x$ is the separation distance along the streamwise direction, and $<\cdot>$ represents the averaging in the temporal and spatially homogeneous directions. The spanwise two-point correlation $R_z$ can be defined in a similar way. 
Fig.~\ref{fig:DQ}($a$) and ($b$) show the variations of the streamwise and spanwise two-point correlations of the above-mentioned variables at $y=0.2h$, respectively. We only show the profiles at $r_x>0$ and $r_z>0$ due to the symmetries of the two correlations with respect to $r_x=0$ and $r_z=0$, respectively.
It can be seen that the profiles of $R_{x}$ and $R_{z}$ of all the quantities overlap with each other well. 
We have checked that the same conclusion applies to other velocity fluctuation components, and the assessment is not shown here for brevity. 
It indicates that these physical quantities share identical length-scale characteristics statistically, at least when $M_b\leq1.5$. 
It also agrees with the observation of \citet{Ringuette2008}, who reported that the correlations of $(\rho u)'$ and $u'$ are very similar in a turbulent boundary layer at Mach 3. 
Whether it holds at higher Mach numbers remains to be checked. Thus, the results presented below mainly focus on the variables weighted by $\sqrt\rho$.

\subsection{Large-scale motions in subsonic and supersonic turbulent channel flows}
\subsubsection{Density-weighted velocity fluctuations}

\begin{figure*}
    \centering
    \includegraphics[width=0.45\linewidth]{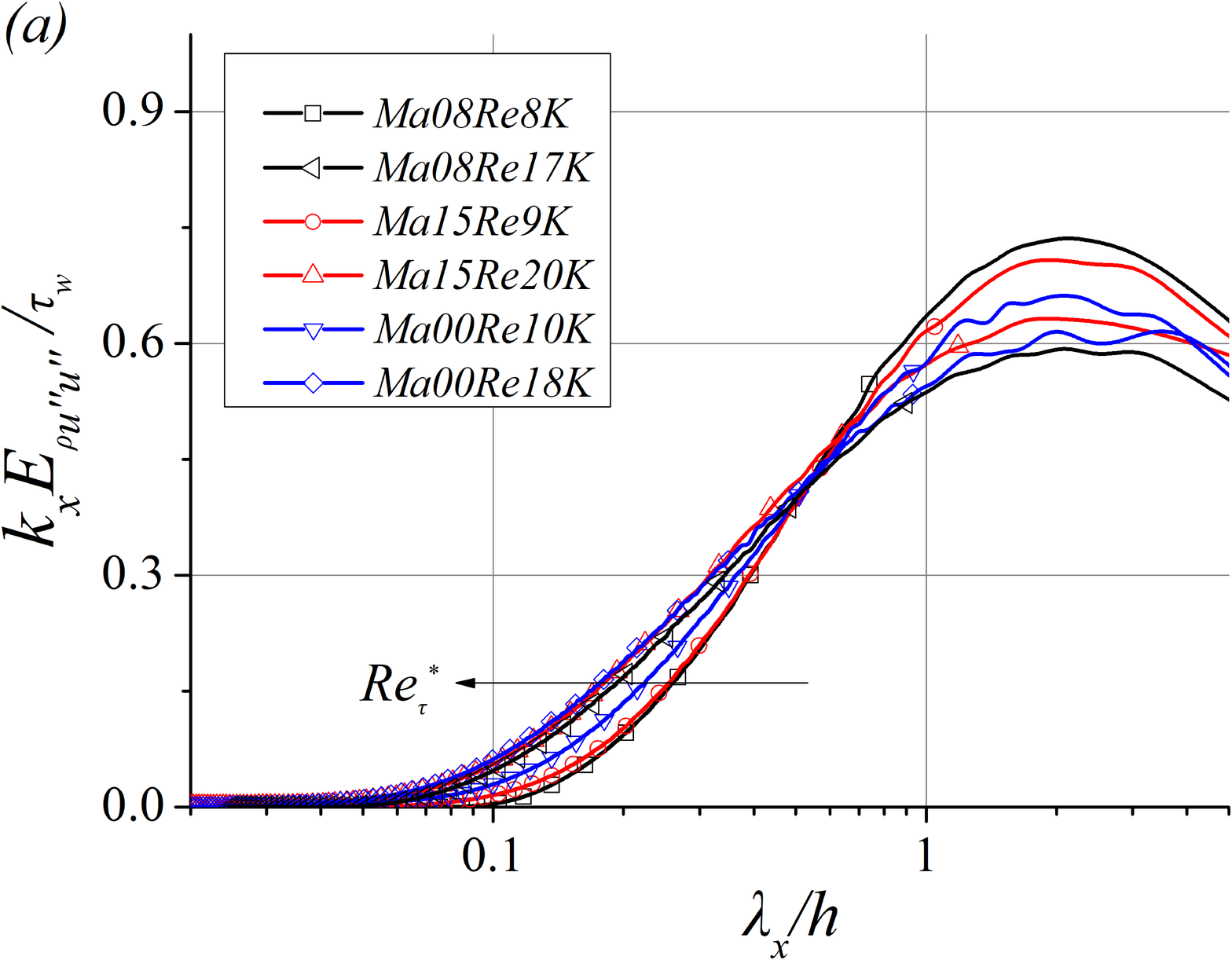}
    \includegraphics[width=0.45\linewidth]{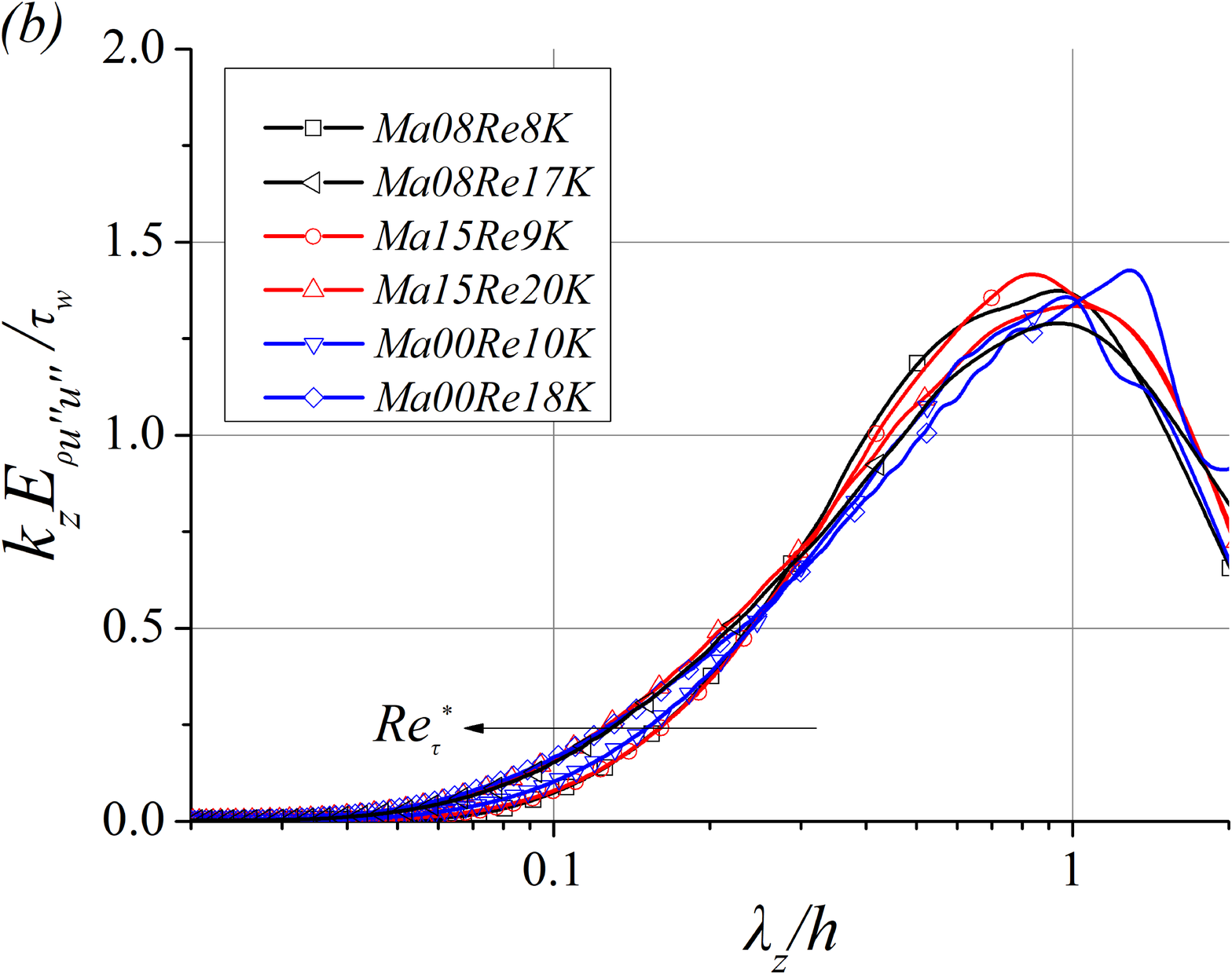}
    \includegraphics[width=0.45\linewidth]{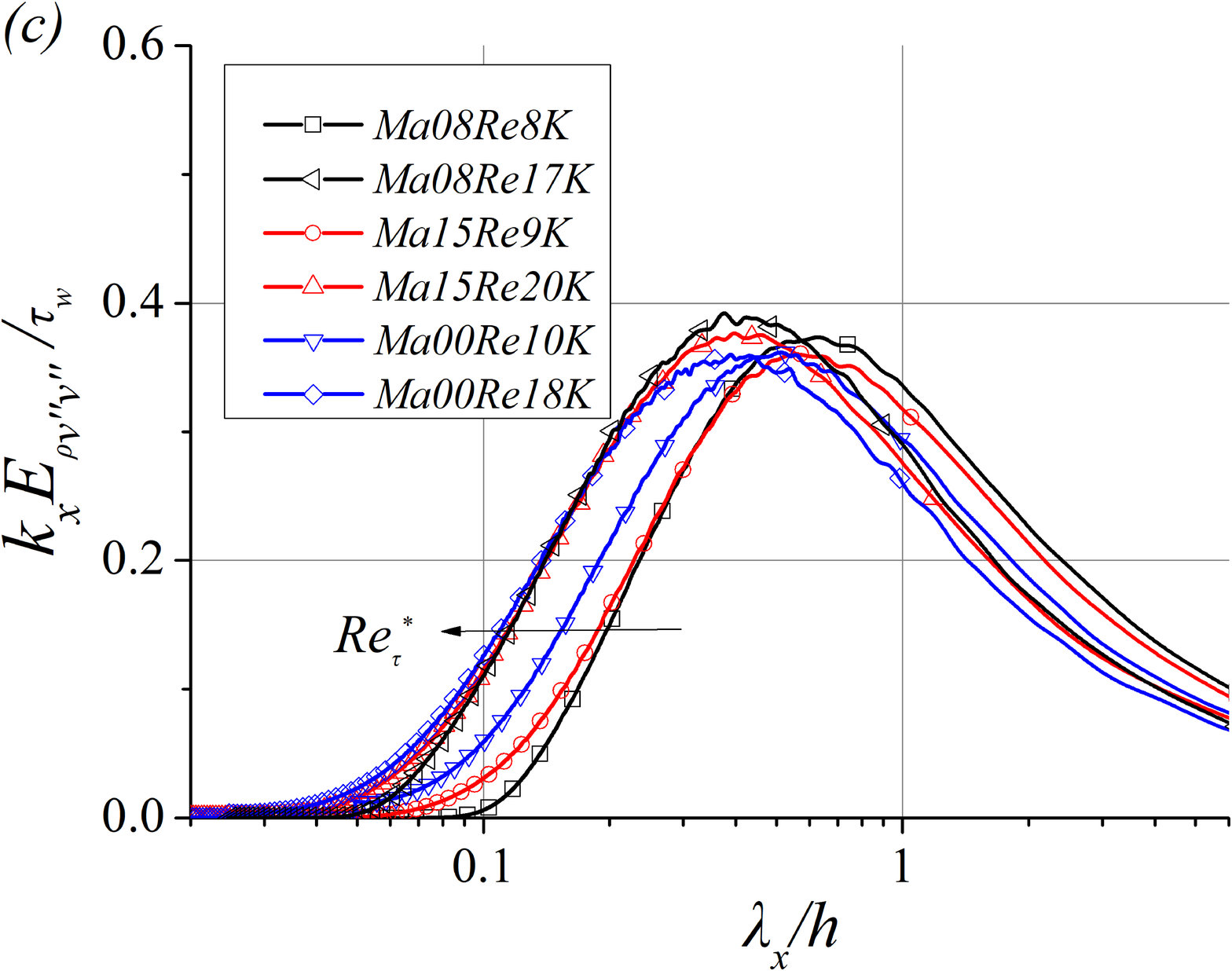}
    \includegraphics[width=0.45\linewidth]{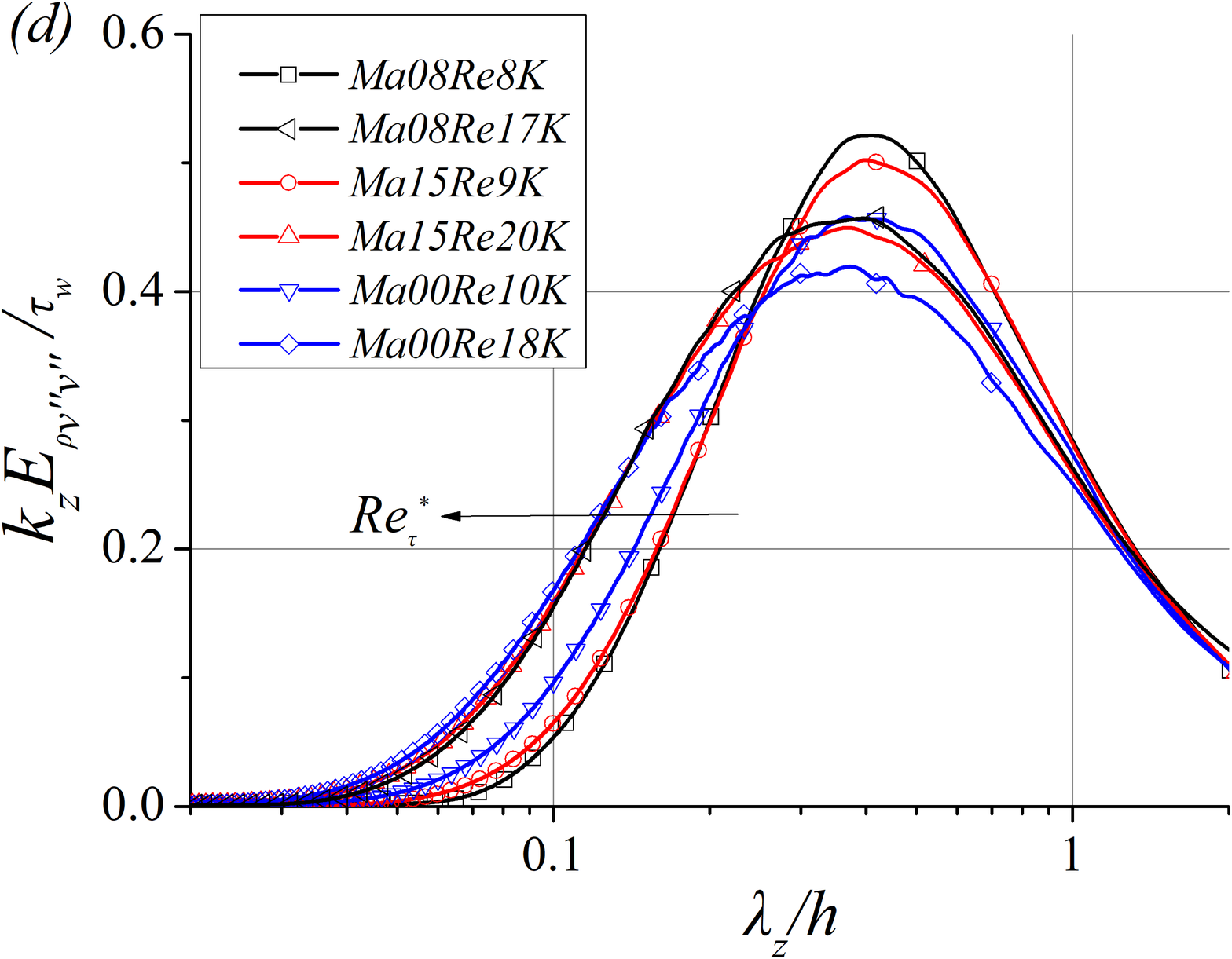}
    \includegraphics[width=0.45\linewidth]{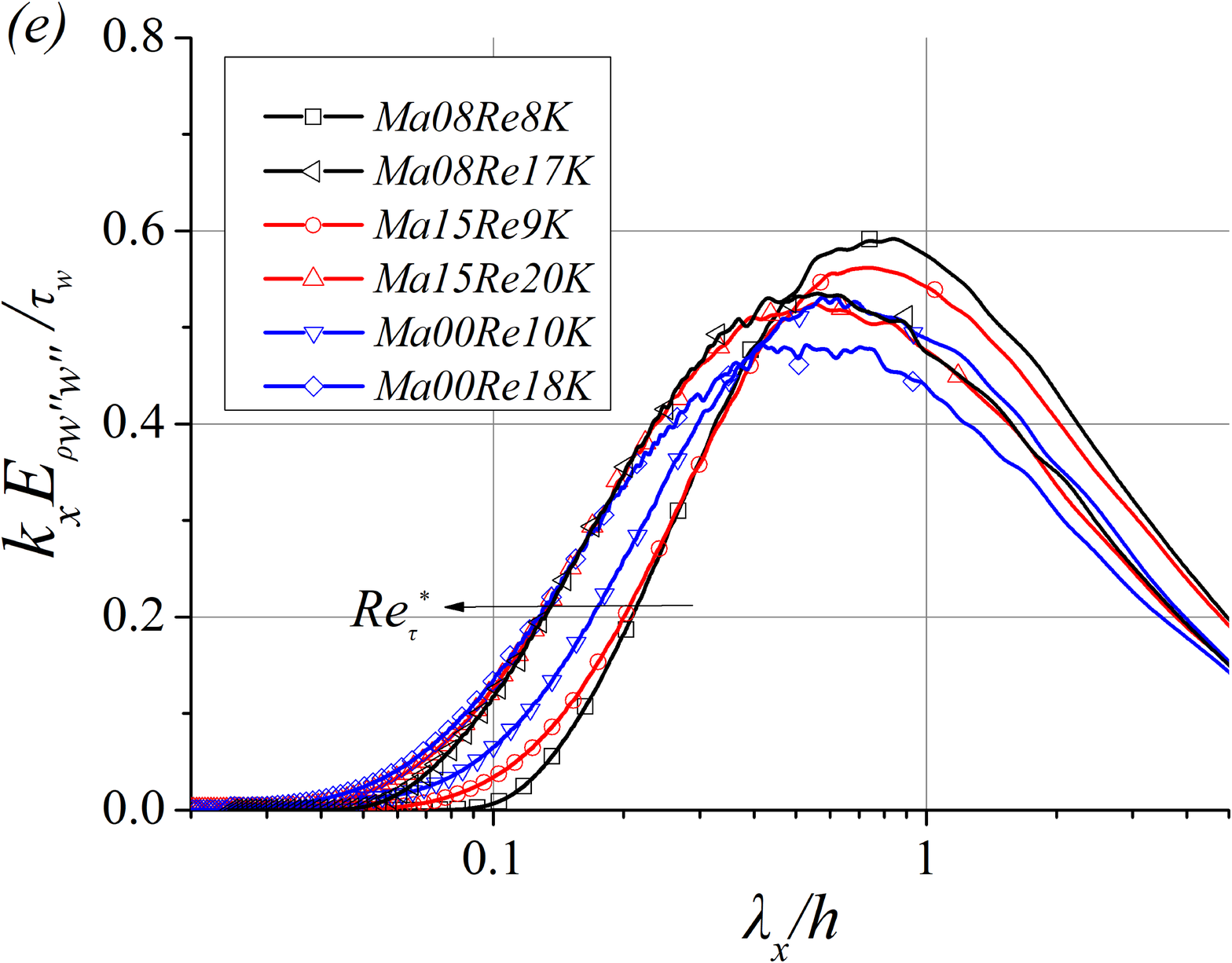}
    \includegraphics[width=0.45\linewidth]{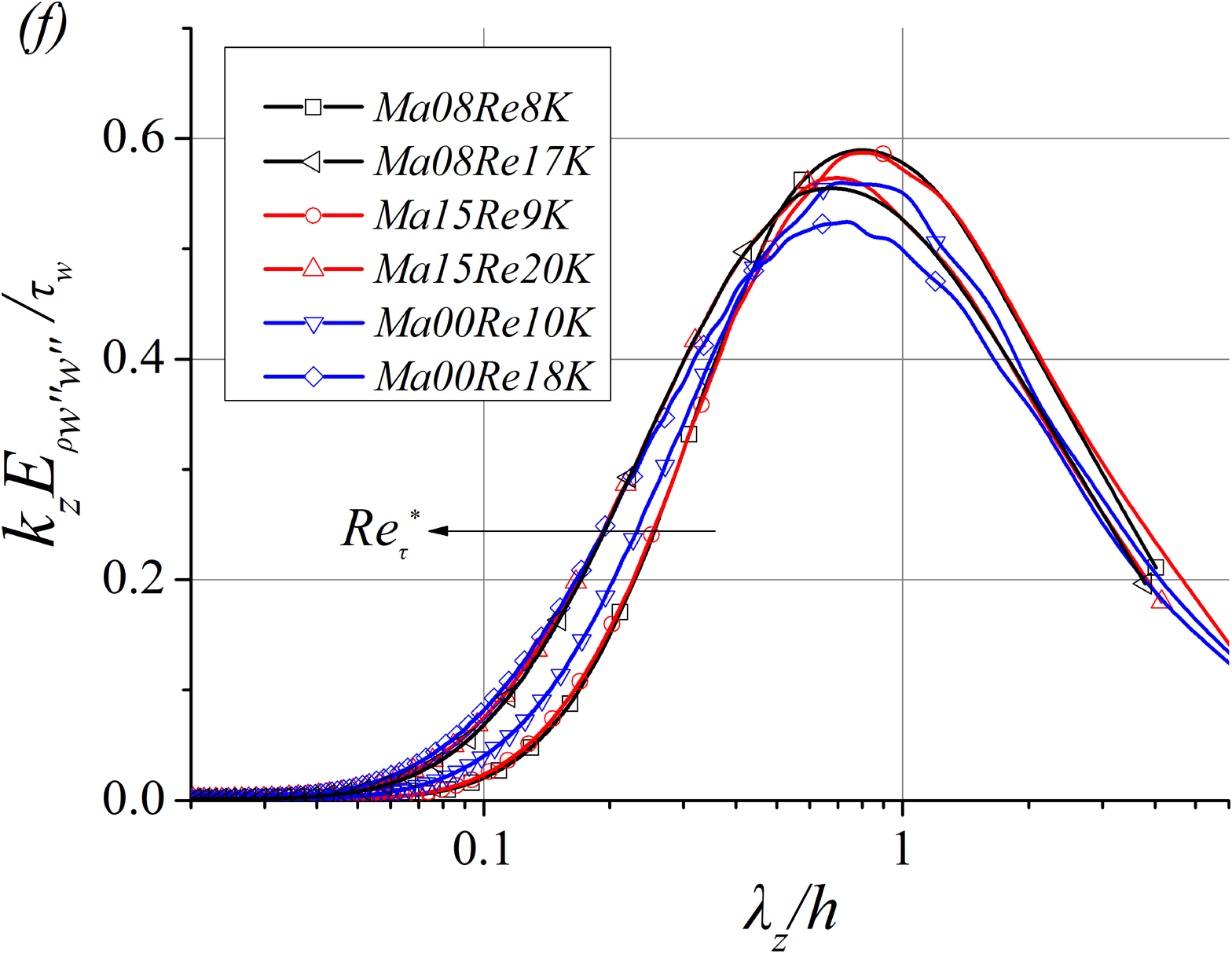}
    \caption{Streamwise ($a$,$c$,$e$) and  spanwise ($b$,$d$,$f$) premultiplied one-dimensional spectra of  $\sqrt\rho u^{''}$ ($a$,$b$), $\sqrt\rho v^{''}$ ($c$,$d$), and $\sqrt\rho w^{''}$ ($e$,$f$) for all cases at $y/h=0.2$.}
    \label{fig:Sp}
\end{figure*}

In this section, we dedicate to comparing the scale characteristics of the large-scale motions in subsonic and supersonic channel flows. Fig.~\ref{fig:Sp}($a$) and ($b$) show the premultiplied streamwise and spanwise spectra of the density-weighted streamwise velocity fluctuations ($\sqrt\rho u^{''}$) at $y=0.2h$. 
The area under the spectral curve represents
the intensity of the density-weighted Reynolds stress $\overline{(\sqrt\rho u^{''})^2}$.
For the streamwise spectra, all the spectra peak at $\lambda_x\approx2h$, and no observable differences in the streamwise wavelengths of the most energetic scales ($\lambda_{x,me}$) between the incompressible, subsonic, and supersonic cases can be found. The spectral energy of the peaks is determined by the magnitude of $Re_{\tau}^*$ instead.
For the spanwise spectra, all the cases peak at $\lambda_z\approx1h$, the typical width of VLSMs \citep{Jimenez2018}, and the distinctions between the $\lambda_{z,me}$ of all cases are still inconspicuous. \textcolor{black}{Similar conclusions about the length scales of $\sqrt\rho u^{''}$ in the outer region have also been drawn by \cite{Yao2020}.}

Very recently, \citet{Bross2021} conducted an experimental study and  observed that the $\lambda_{x,me}$ in the outer region of the supersonic turbulent boundary layers are slightly larger than those of the subsonic cases, whereas the $\lambda_{z,me}$ of the supersonic cases have a distinct increase compared with the subsonic cases. 
In compressible turbulent channel flows, at least within the Reynolds and Mach numbers under scrutinizing by the present study, this phenomenon is not that remarkable. 
The possible explanations are given in section \ref{DIS}.
As a side note, \citet{Bross2021} normalized the spectra of the streamwise velocity fluctuations  in a different way, i.e., $(\overline{\rho}/\rho_w)k_xE_{u'u'}/u_{\tau}^2$ and $(\overline{\rho}/\rho_w)k_zE_{u'u'}/u_{\tau}^2$. We have verified that the conclusions reached above are unaffected by normalizing the spectra in this manner (see section \ref{DIS}). 

The premultiplied streamwise and spanwise spectra of $\sqrt\rho v^{''}$ and $\sqrt\rho w^{''}$ are also shown in
Fig.~\ref{fig:Sp}($c$-$d$) and ($e$-$f$), respectively. 
It can be seen that no clear Mach-number dependence of the wavelengths of the spectral peaks can be identified for all cases, regardless of the spectral direction.
Moreover, the spectra of the cases at larger $Re_{\tau}^*$ are more energetic in small length scales, and the cases at similar $Re_{\tau}^*$ collapse well. 
Although less obvious, this phenomenon can also be seen in the premultiplied streamwise and spanwise spectra of $\sqrt\rho u^{''}$. 
These observations highlight the fact that the semilocal Reynolds-number effects dominate the energy distribution among the multi-scale eddies in the outer region, regardless of whether the flow passes the sound barrier or not, and whether the flow is compressible or not.
They are also consistent with some previous studies \citep{Trettel2016,Patel2016,Hirai2021,Griffin2021,Bai2022compressible}, which emphasize that the semilocal Reynolds number $Re_{\tau}^*$ is a key similarity parameter for the compressible wall-bounded turbulence. 
In summary, the present study does not observe that there is a sudden increase in length scales of the out-layer motions when the flow transitions from subsonic to supersonic state.

\subsubsection{Thermodynamic variables}
\begin{figure*}
    \centering
    \includegraphics[width=0.45\linewidth]{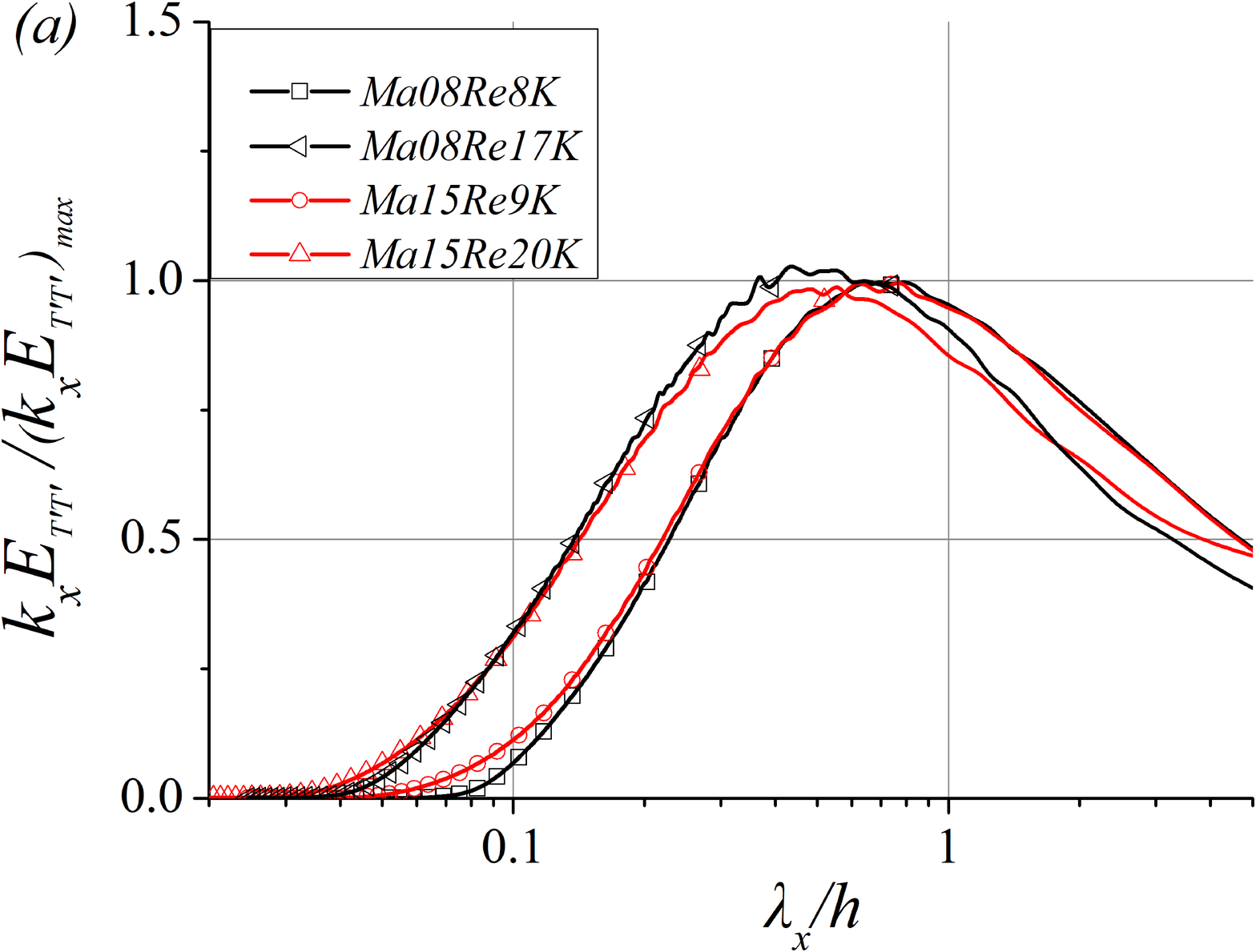}
    \includegraphics[width=0.45\linewidth]{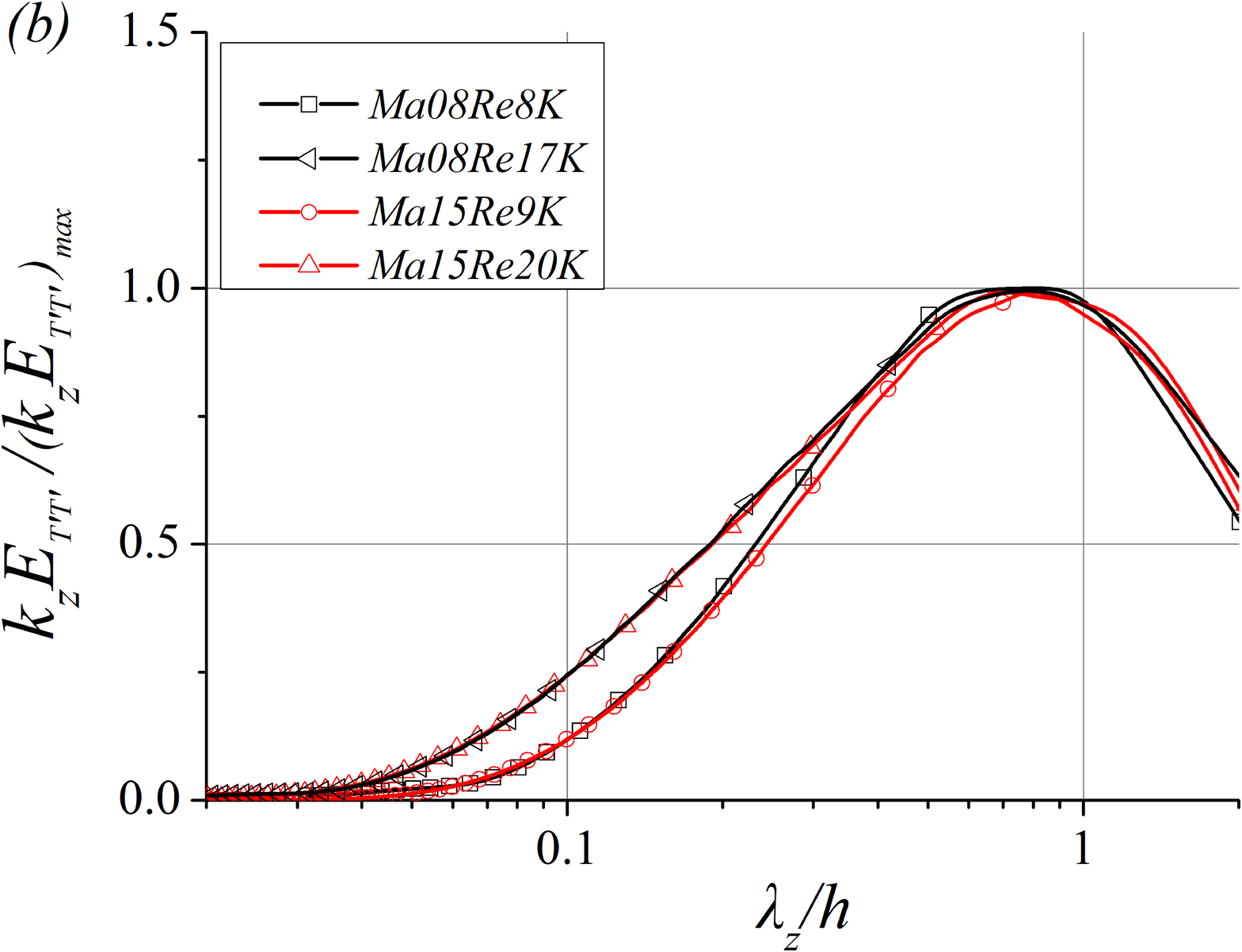}
    \includegraphics[width=0.45\linewidth]{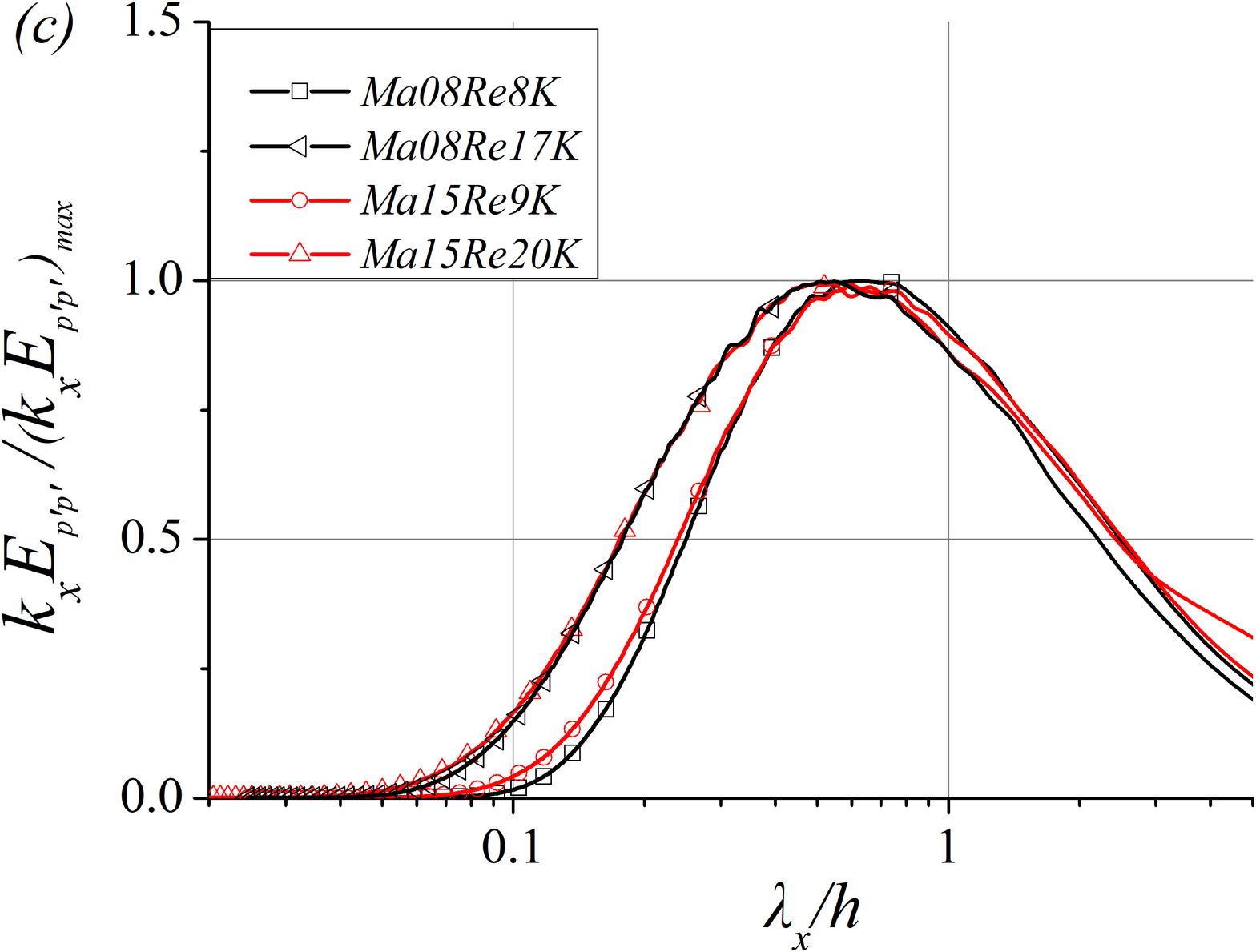}
    \includegraphics[width=0.45\linewidth]{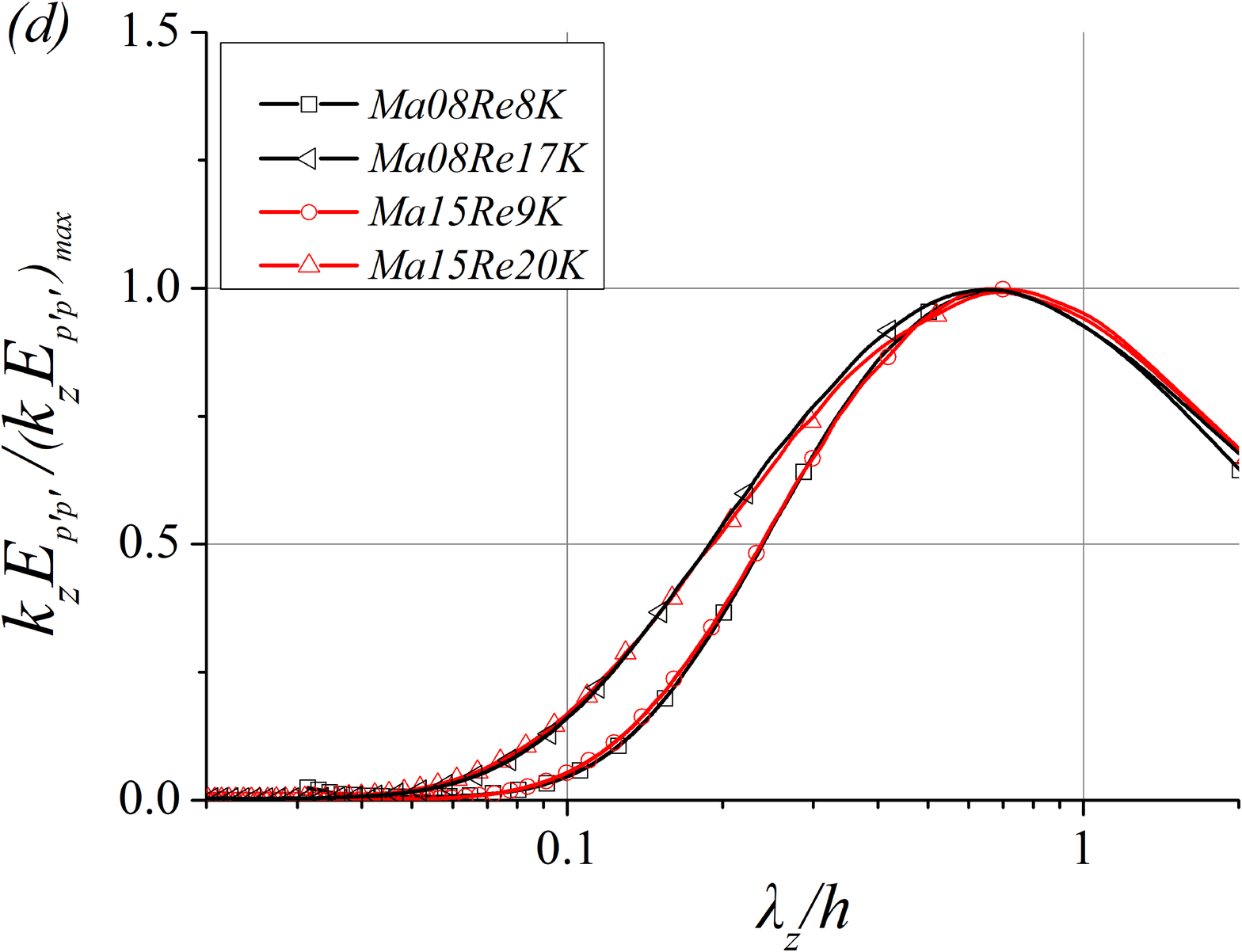}
    \includegraphics[width=0.45\linewidth]{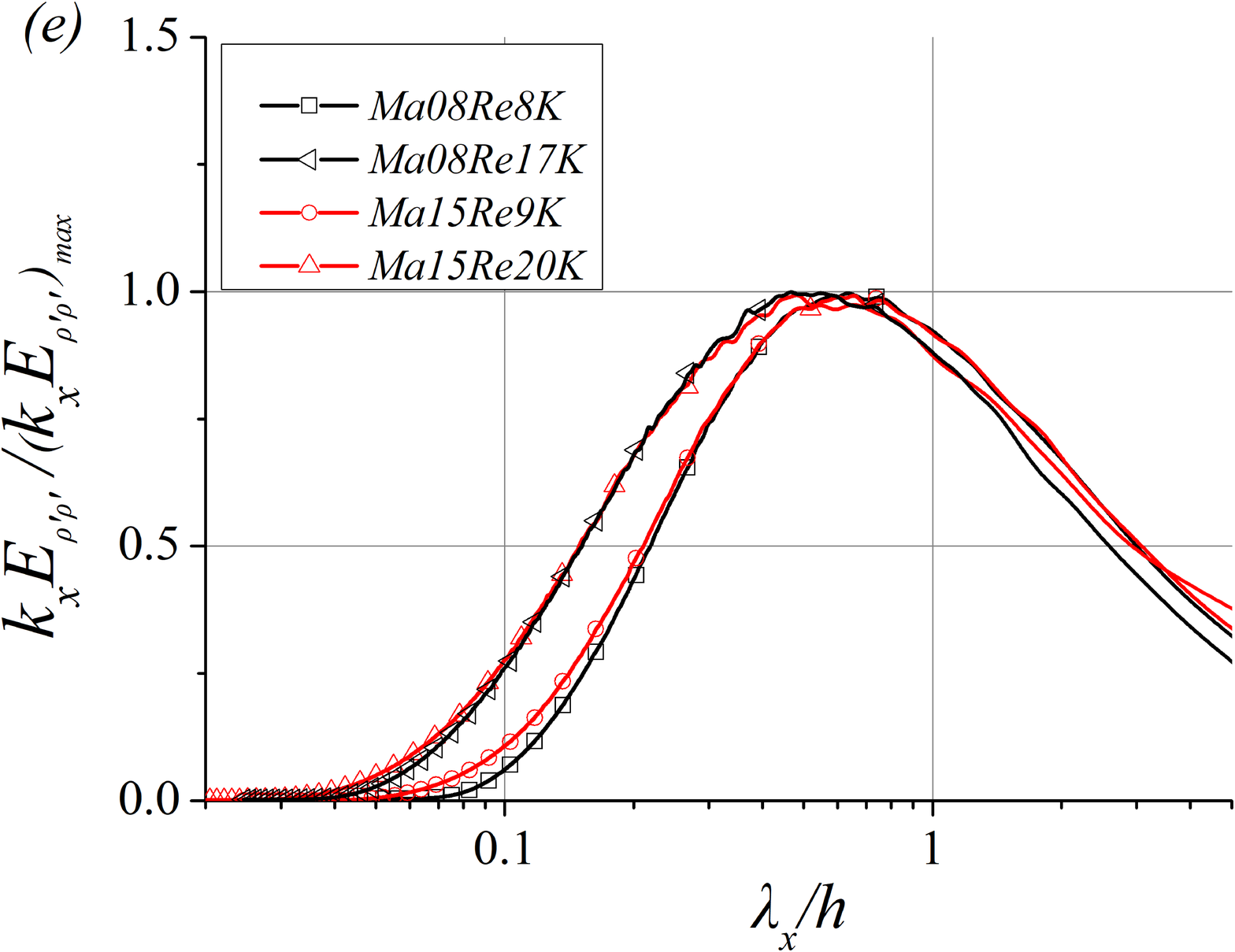}
    \includegraphics[width=0.45\linewidth]{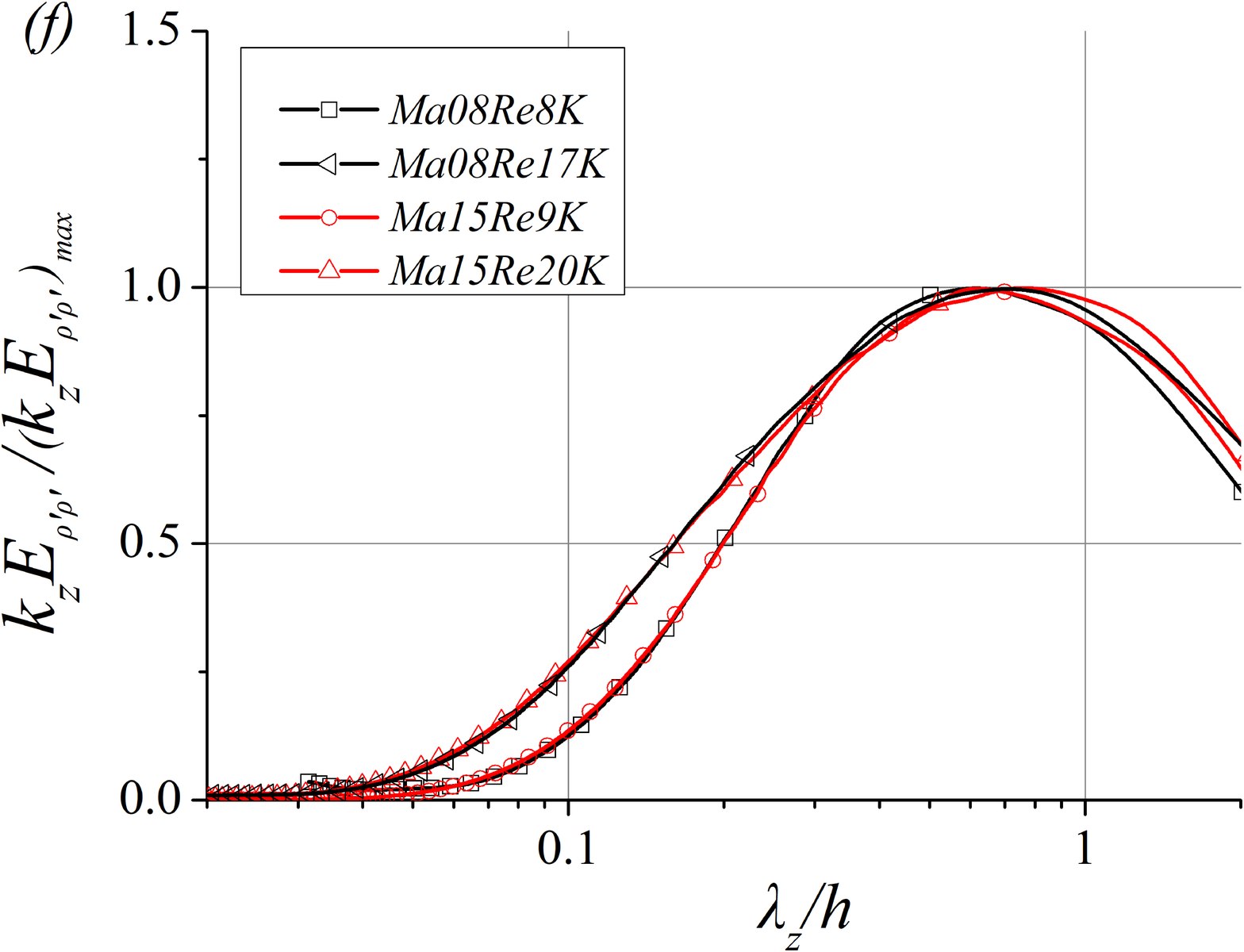}
    \caption{Streamwise ($a$,$c$,$e$) and  spanwise ($b$,$d$,$f$) premultiplied one-dimensional spectra of  $T'$ ($a$,$b$), $p'$ ($c$,$d$), and $\rho'$ ($e$,$f$) for all compressible cases at $y/h=0.2$. All the spectra are normalized by their maximum values.}
    \label{fig:Sp2}
\end{figure*}
In this section, we turn to report the scale characteristics of the thermodynamic variables, i.e., $T'$, $p'$, and $\rho'$ at outer region $y=0.2h$. 
Fig.~\ref{fig:Sp2}($a$-$b$), ($c$-$d$), and ($e$-$f$) show the premultiplied streamwise and spanwise spectra of $T'$, $p'$, and $\rho'$ of the compressible cases, respectively. 
All the spectra are normalized by their maximum values for comparison. 
It is apparent that the cases with similar  $Re_{\tau}^*$ share akin scale-dependent energy distributions for all three thermodynamic variables. 
This observation supports the viewpoint put forward above that $Re_{\tau}^*$ is the key parameter in shaping the spectra.
Moreover, no clear Mach-number effects on the scale characteristics can be identified between the cases with similar $Re_{\tau}^*$. 
This demonstrates once again that there is no sudden increase in length scales of the outer-region motions when the flow becomes supersonic from a subsonic state, as the scale characteristics of the thermodynamic variables are highly linked with those of LSMs and VLSMs \citep{Pirozzoli2011}.

\subsubsection{Discussions}\label{DIS}

\begin{figure*}
    \centering
    \includegraphics[width=0.45\linewidth]{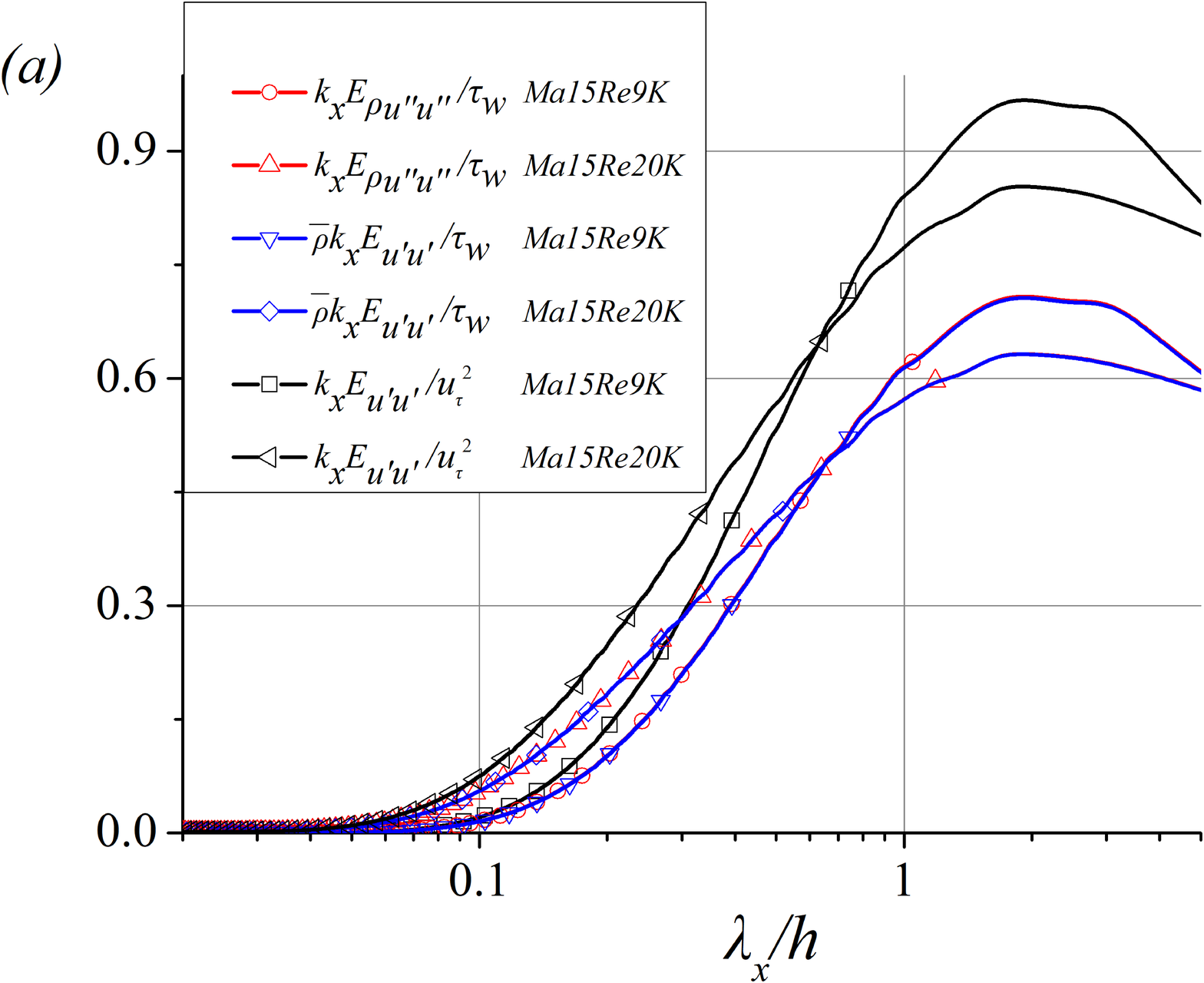}
    \includegraphics[width=0.45\linewidth]{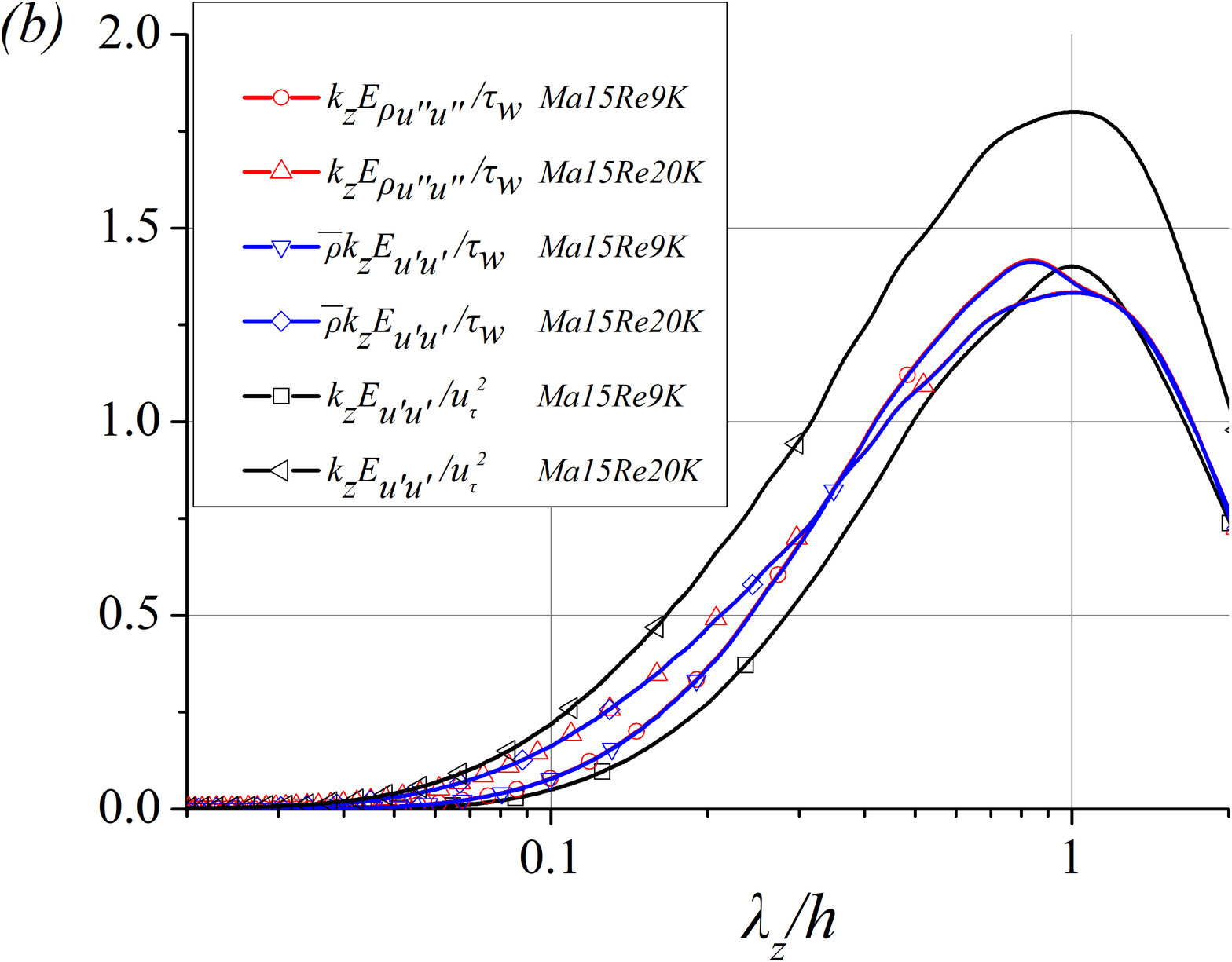}
    \caption{Streamwise ($a$) and spanwise ($b$) premultiplied spectra of the streamwise velocity fluctuations for Ma15Re9K and Ma15Re20K at $y=0.2h$ with different density-weighted approaches and without density weighting.}
    \label{fig:Sp3}
\end{figure*}

The reason why the DNS results differ from the experimental results merits a discussion.
\citet{Ganapathisubramani2006} and \citet{Bross2021} conjectured that analysis of mass-flux $(\rho u)'$ versus $u'$ may result in distinct conclusions about the length scales of the outer motions. 
The origin of this discrepancy is ascribed to the different
velocity measurement tools in the experiments, i.e., the hot wire versus the PIV. 
However, Fig.~\ref{fig:DQ} above demonstrates that the discrepancies of the mean length scale between these two variables at $y=0.2h$ are minor, at least, within the Reynolds and Mach numbers under consideration.
Fig.~\ref{fig:Sp3} compares the streamwise and spanwise spectra of the streamwise velocity fluctuations for Ma15Re9K and Ma15Re20K with different density-weighting approaches and without density weighting. 
It is noted that $(\overline{\rho}/\rho_w)k_xE_{u'u'}/u_{\tau}^2$ and $(\overline{\rho}/\rho_w)k_zE_{u'u'}/u_{\tau}^2$ are commonly adopted by experimental studies to exhibit the scale characteristics of the velocity signals. 
It can be seen that the density-weighted spectra overlap with each other. 
It indicates that the different weighting methods do not alter the energy distributions among scales within the Reynolds and Mach numbers under consideration. 
In addition, though the spectra without density weighting have unique forms, the length scales of their peaks are consistent with the density-weighted ones, that is, $\lambda_{x,e}\approx2h$ and $\lambda_{z,e}\approx1h$. 
In other words, for the spectra of the current examples under examination, the consequences of weighting with density on the values of $\lambda_{x,e}$ and $\lambda_{z,e}$ are negligible. 

Another possible reason is that the configuration of the wall turbulence here is different from that of the work of experimentalists, who usually investigate the length scales of outer motions in turbulent boundary layers, instead of channel flow \citep{Smits1989,Ganapathisubramani2006,Elsinga2010,Bross2021}. 
Some previous studies on incompressible wall turbulence have reported that the average streamwise length scale of VLSMs in internal flow (e.g., pipe and channel flow) is larger than that of the turbulent boundary layer  \citep{Monty2009,Lee2013,Sillero2014}. 
Besides, the friction Reynolds numbers of the cases in the present study are $O$($10^2$)
to $O$($10^3$), whereas for the works of experimentalists, they are usually $O$($10^3$) to $O$($10^4$) \citep{Ganapathisubramani2006,Bross2021}. Hence, the observations reported in the present study remained to be verified in high-Reynolds-number wall turbulence.
We also notice that, in the experiments conducted by \citet{Bross2021}, the walls are hot (the wall temperature is higher than that of free stream), whereas for the turbulent channel flows in the present study, the walls are cool (see Fig.~\ref{fig:valid1} and \ref{fig:valid2}). This might be another non-negligible factor.

\begin{figure*}
    \centering
    \includegraphics[width=0.5\linewidth]{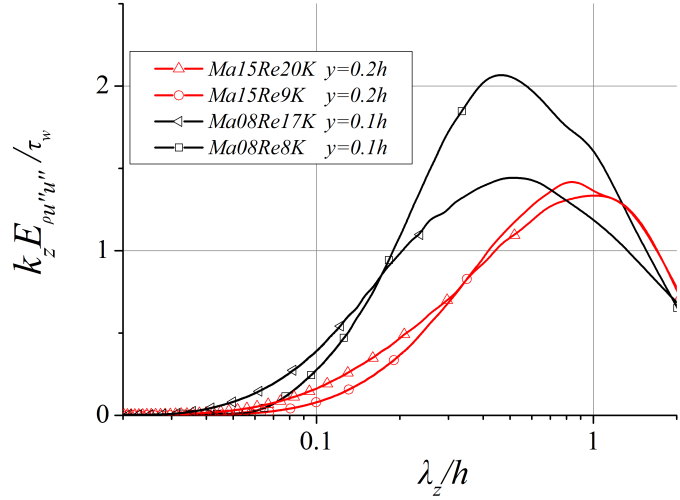}
    \caption{Spanwise premultiplied spectra of $\sqrt\rho u^{''}$ at $y=0.1h$ for the subsonic cases, and at $y=0.2h$ for the supersonic cases.}
    \label{fig:Sp4}
\end{figure*}

At last, the wall-normal location of the measurement plane deserves a discussion. 
When examining the spanwise length scales of the outer motions, \citet{Bross2021} set the measurement plane at approximately $y/\delta_{}=0.1$ for the subsonic cases, and at $y/\delta_{}=0.2$ for the supersonic cases.
They found that the average width of the outer motions is larger for the supersonic cases. 
In Fig.~\ref{fig:Sp4}, we repeat this process, and compare the spanwise spectra of $\sqrt\rho u^{''}$ for different cases. A similar phenomenon can also be observed. 
However, as the spanwise length scales
of the energy-containing motions populating the logarithmic and outer regions scale in the outer unit and
increase with their wall-normal heights \citep{DelAlamo2004,Hwang2015,Cheng2019}, the discrepancies of  $\lambda_{z,e}$ for the supersonic and the subsonic cases shown in Fig.~\ref{fig:Sp4} here and Fig.~15 of \citep{Bross2021} are ascribed to the wall-normal growth of the motions, rather than the Mach-number effects. 
Thus, to preclude the scale-growth effects and uncover the Mach-number effects on length scales of LSMs and VLSMs, the wall-normal position of the measurement planes for the supersonic and the subsonic cases should be identical and located in the logarithmic or outer regions.

\subsection{Self-similar structures in compressible turbulent channel flows}
In this section, we pay attention to the self-similar structures in compressible turbulent channel flows. 
To inspect the structures that are highly correlated with the near-wall flows, the linear coherence spectrum (LCS) of the variable $\psi$ is adopted, which is given by
\begin{equation}
\gamma^{2}\left(y_o, y ; \lambda_{x}\right) \equiv \frac{\left|\left\langle\widehat{\psi}\left(y_o ; \lambda_{x}\right) \widehat{\psi}^*\left(y ; \lambda_{x}\right)\right\rangle\right|^{2}}{\left\langle\left|\widehat{\psi}\left(y_o ; \lambda_{x}\right)\right|^{2}\right\rangle\left\langle\left|\widehat{\psi}\left(y ; \lambda_{x}\right)\right|^{2}\right\rangle},
\end{equation}
where  $\psi(y_o)$ denotes the variable $\psi$ at the location $y_o$ in the logarithmic region, $\psi(y)$ denotes the variable $\psi$ at the location $y$ in the near-wall region, $\widehat{\psi}$ is the Fourier coefficient of $\psi$, $\widehat{\psi}^{*}$ is the complex conjugate of $\widehat{\psi}$, and $\mid \cdot \mid$ is the modulus. 
$\gamma^{2}$ evaluates the square of the scale-specific correlation between $\psi(y_o)$ and $\psi(y)$  with $0\leq\gamma^{2}\leq1$ \citep{Bendat2011}. 
LCS has been adopted to study the self-similar wall-attached structures in incompressible boundary layers \citep{Baars2017,Baars2020a}\textcolor{black}{, and the coherence between the temperature and the velocity fields in compressible wall turbulence \citep{Yu2021}.} 
To the authors' knowledge, no study on the \textcolor{black}{geometrical characteristics of the} self-similar structures in compressible turbulent channel flows by employing  LCS has been reported, particularly the structures of the thermodynamic variables. 
In the present study, we fill this gap via analyzing the cases Ma08Re17K and Ma15Re20K, as their Reynolds numbers are  relatively higher.
\begin{figure*}
    \centering
    \includegraphics[width=0.45\linewidth]{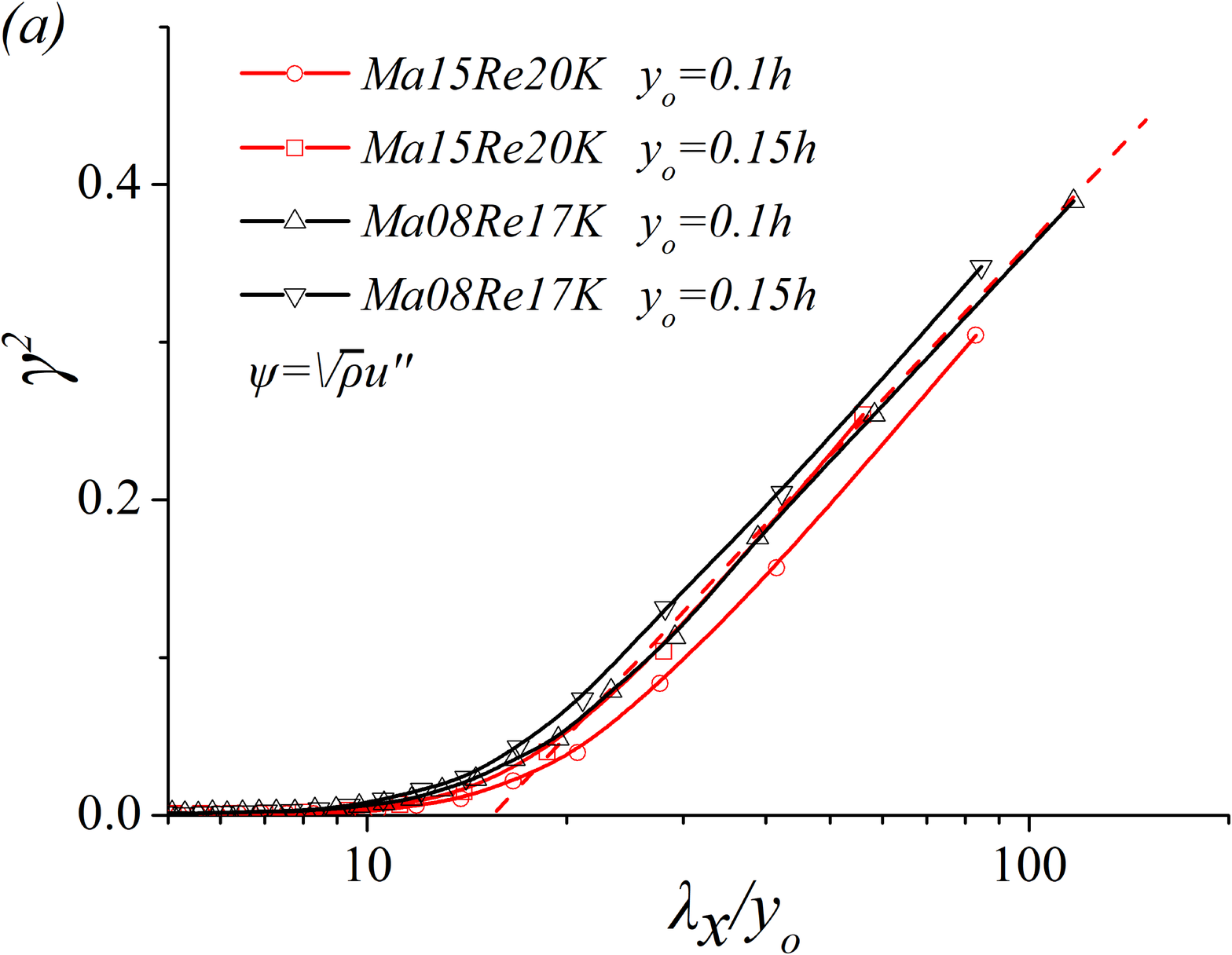}
    \includegraphics[width=0.45\linewidth]{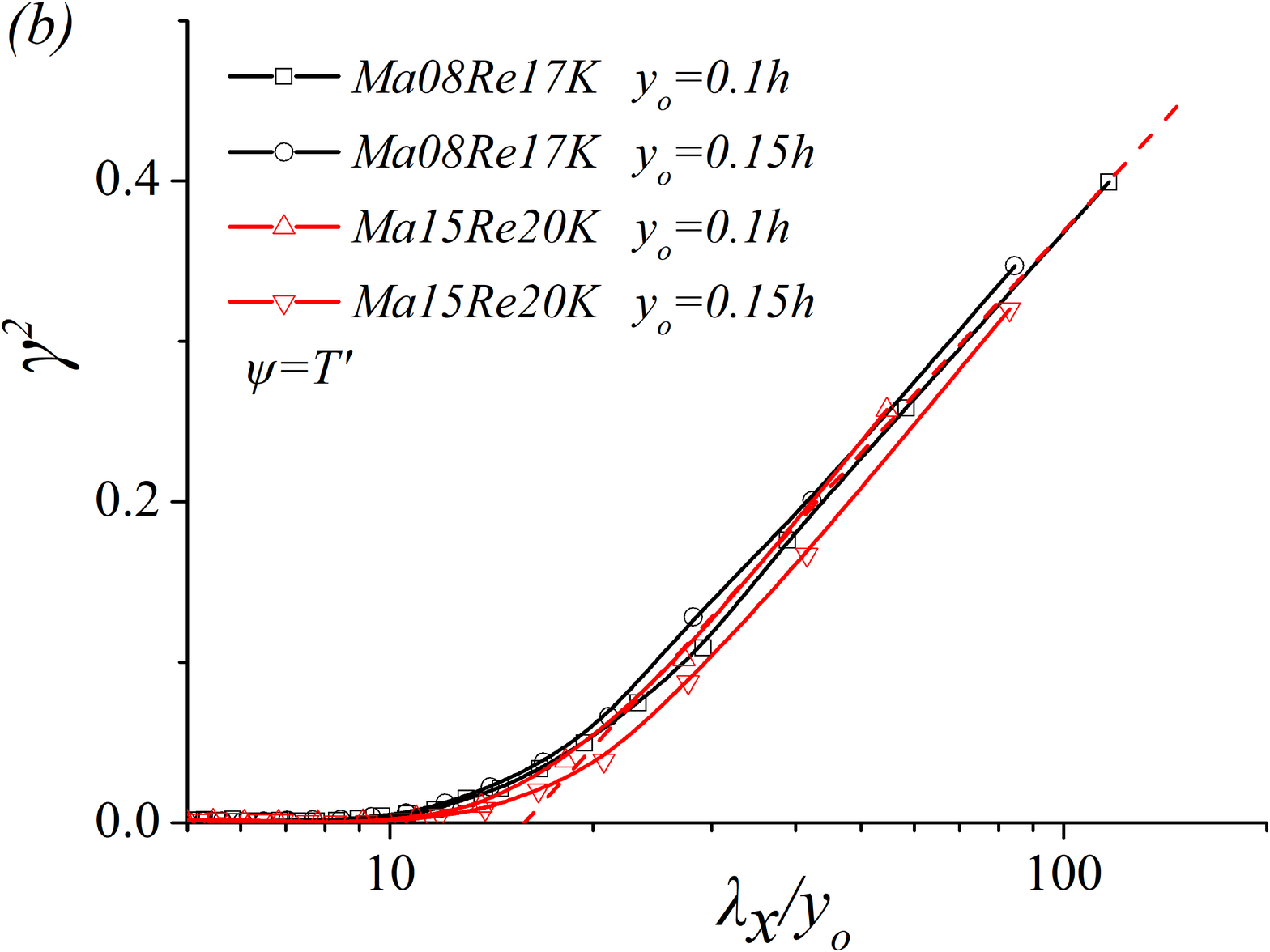}
    \includegraphics[width=0.45\linewidth]{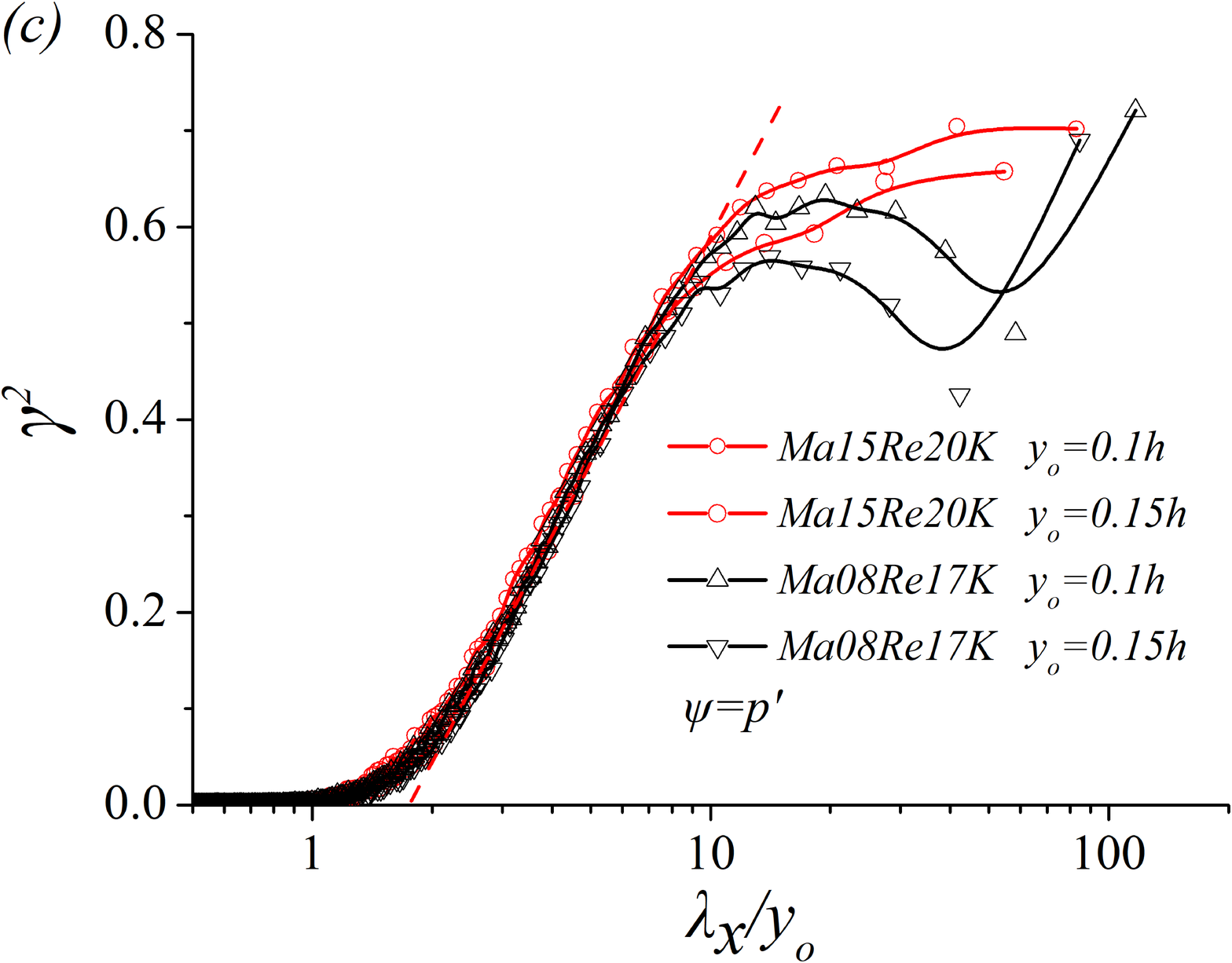}
     \includegraphics[width=0.45\linewidth]{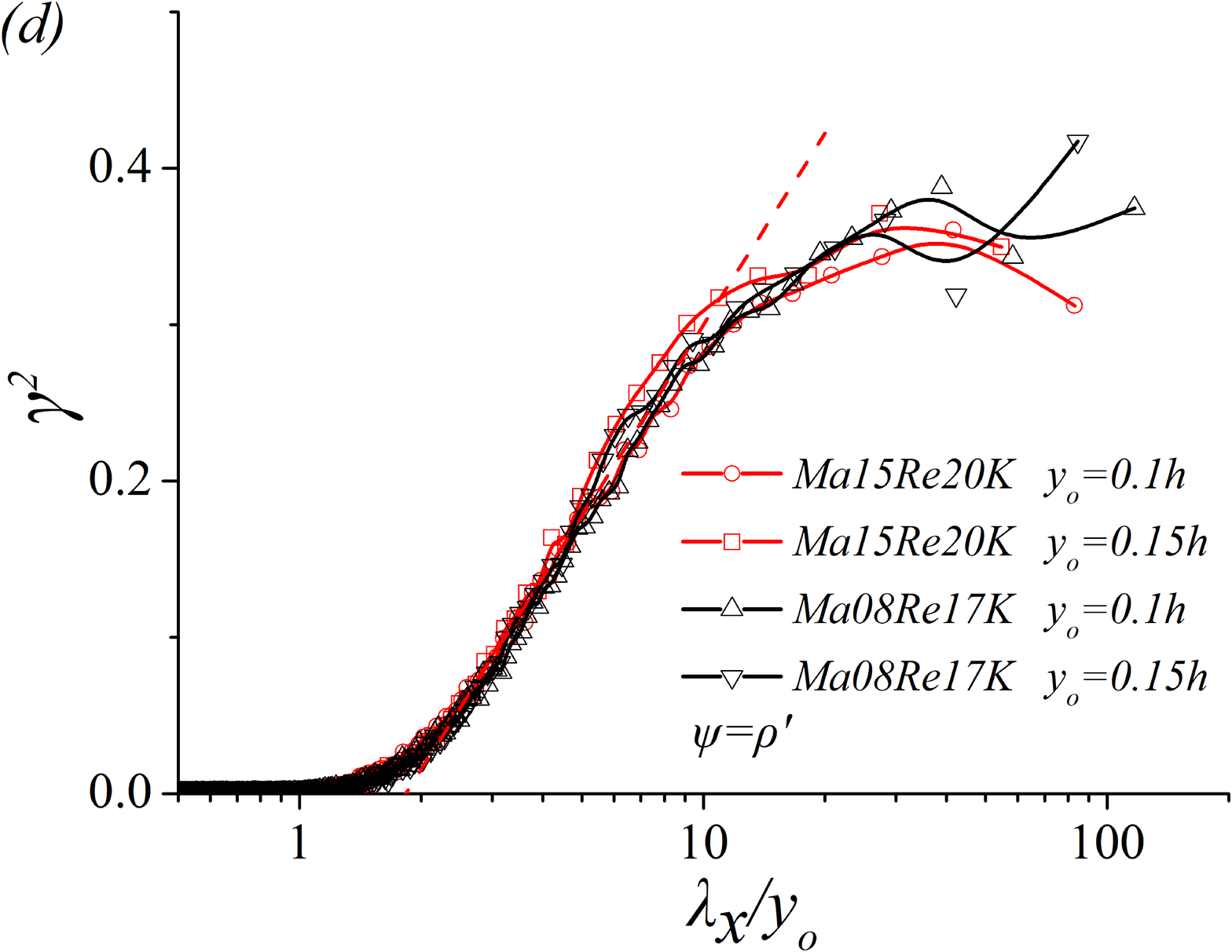}
    \caption{Variations of $\gamma^{2}$ as
    functions of $\lambda_x/y_o$ at two selected wall-normal heights, i.e., $y_o=0.1h$ and $y_o=0.15h$, for $\psi=\sqrt\rho u^{''}$ ($a$), $T'$ ($b$), $p'$ ($c$), and $\rho'$ ($d$), respectively. The dashed lines
    in the subfigures denote the logarithmic growth Eq.~(\ref{gama}).}
    \label{fig:LCS}
\end{figure*}

In the present study, the near-wall region is fixed at $y^+\approx0.6$, and the outer reference height
$y_o$ is located in the logarithmic region. 
We have checked that as long as  $y^+\le 15$, the results presented below are insensitive to the choice of $y^+$. 
Fig.~\ref{fig:LCS}($a$-$d$) show the variations of $\gamma^{2}$ as
functions of $\lambda_x/y_o$ at two selected wall-normal heights, i.e., $y_o=0.1$ and $y_o=0.15$, for $\psi=\sqrt\rho u^{''}$, $T'$, $p'$, and $\rho'$, respectively. 
For $\psi=\sqrt\rho u^{''}$ and $T'$, $\gamma^{2}$ grows linearly with $\ln(\lambda_x/y_o)$ for $\lambda_x/y_o\ge20$, whereas for $\psi=p'$ and $\rho'$, the similar linear relations can be recognized for $2\le\lambda_x/y_o\le10$. 
Moreover, for the wall turbulence at different wall-normal heights and Mach numbers, the $\gamma^{2}$ profiles of a target variable $\psi$ roughly coincide with each other in these regions. 
It indicates that there are self-similar structures of $\psi$ populating the logarithmic region, whose scale characteristics are not sensitive to the Mach numbers. 
This scenario is consistent with the celebrated attached-eddy model \citep{Townsend1976,Perry1982}, which hypothesizes that the energy-containing motions in the logarithmic region are self-similar and can permeate into the near-wall region. 

\begin{table}[!htbp]
\centering
\begin{tabular}{cccc}
\hline
\hline
$\psi$   & $C_1$ & $C_2$ &  $AR$ \\ 
\hline
\hline
$\sqrt\rho u^{''}$  & 0.194& -0.531 & 15.44 \\
\hline
$T'$ & 0.20& -0.552 & 15.79 \\
\hline
$p'$ & 0.339& -0.190 & 1.75 \\
\hline
$\rho'$ & 0.177& -0.107 & 1.83 \\
\hline
\end{tabular}
\caption{The fitted constants $C_1$, $C_2$ in Eq.~(\ref{gama}), and the corresponding $AR$ assessed by Eq.~(\ref{gama2}) for different $\psi$. \label{tab:AR}}
\end{table}

The magnitude of $\gamma^{2}$ within the self-similar region takes the form of
\begin{equation}\label{gama}
\gamma^{2}=C_{1} \ln \left(\frac{\lambda_{x}}{y_o}\right)+C_{2},
\end{equation}
where $C_1$ and $C_2$ are two constants and can be estimated by fitting the $\gamma^{2}-$ spectra over this range. \citet{Baars2017} proposed that the streamwise/wall-normal aspect ratio of the self-similar structures can be assessed by
\begin{equation}\label{gama2}
\left.A R \equiv \frac{\lambda_{x}}{y_o}\right|_{\gamma^{2}=0}=\exp \left(\frac{-C_{2}}{C_{1}}\right),
\end{equation}
according to the hierarchical structure of the self-similar attached eddies \citep{Perry1982}. 
The values of $C_1$, $C_2$, and 
$AR$ for distinct $\psi$ are listed in Table \ref{tab:AR}. 
It can be seen that the $AR$ of $\sqrt\rho u^{''}$ and $T'$ are very close to each other and appropriately equal to 15.5. 
This result is in accordance with the assessment conducted by \citet{Baars2017}, who found the $AR$ of the self-similar structures of the streamwise velocity fluctuations in incompressible turbulent boundary layers is 14. 
Besides, it is not unexpected that
the $AR$ of $T'$ is closely approximate to that of $\sqrt\rho u^{''}$, as a number of studies show that the strong Reynolds analogy (SRA) is valid in compressible wall turbulence \citep{Morkovin1962,Huang1995,Cebeci2012,fu2021shock}. 
For $p'$ and $\rho'$, the $AR$ is approximately equal to 1.8, and significantly smaller than those of $\sqrt\rho u^{''}$ and $T'$. 
This observation is reminiscent of the property of $p'$ in incompressible wall turbulence, that is, $p'$ is more energetic at small scales ($\lambda_x \le C_3h$, $C_3$ is a constant) \citep{Tsuji2016}. 
\citet{Baars2017} and \citet{Baars2020a} also noticed that the $\gamma^{2}-$spectra of $u'$ in incompressible turbulent boundary layers tend to be constant at an outer-scaling limit of $\lambda_x/h\approx10$. 
It can not be observed by the present study for  $\psi=\sqrt\rho u^{''}$ and $T'$ due to the limited size of the computational domain. 
However, for $\psi=p'$ and $\rho'$, the $\gamma^{2}-$spectra shown in Fig.~\ref{fig:LCS}($c$) and ($d$) departure from the power-law relation Eq.~(\ref{gama}) for $\lambda_x/y_0\ge10$, which results from the very large-scale structures of these two variables. 
Again, this shows that the scale characteristics of $p'$ and $\rho'$ are completely different from those of
$\sqrt\rho u^{''}$ and $T'$ in compressible wall turbulence.

\textcolor{black}{Finally, we want to point out that the geometrical characteristics of the self-similar structures reported in the present study may aid in developing physics
-based data-driven filters in compressible wall turbulence study, just as the work of \cite{Baars2020a} in incompressible flows. Besides, the LCS employed here can also be used to shed light on the effects of the wall thermal boundary conditions on the sizes of the self-similar structures in the compressible flows. For example, \citet{Huang2022} observed that strong wall cooling can lead to a remarkable reduction in the scale separation between the small and large eddies. The LCS can be developed as an effective tool to quantify this variation.}

\section{Conclusions}\label{CO}
In the present study, we conduct a series of DNSs of the subsonic and supersonic turbulent channel flows
at moderate Reynolds numbers. By employing the DNS database, we compare the scale characteristics of the outer motions in subsonic and supersonic turbulence and dissect the geometrical characteristics of the self-similar structures for the streamwise velocity fluctuations and the fluctuations of the thermodynamic variables. The conclusions are summarized below.

($a$) The semilocal friction Reynolds number rather than the Mach number determines the energy distribution among the multi-scale structures in the outer region, not only for the velocity fluctuations but also for the fluctuations of the thermodynamic variables. In addition, contrary to what a prior experimental study claimed, we have not found that the streamwise and spanwise length scales of the outer motions alter significantly when flow passes the sound barrier. 

($b$) The streamwise/wall-normal aspect ratio of the self-similar wall-attached structures of $\sqrt\rho u^{''}$ and $T'$ is $15.5$, which is similar to that of the self-similar structures in incompressible wall turbulence. The counterpart of $p'$ and $\rho'$ is $1.8$, which has not been reported before. The very large-scale structures of $p'$ and $\rho'$ are those with $\lambda_x/y_0\ge10$. These scale characteristics are found to be not sensitive to the Mach and Reynolds numbers.

It is noted that the conclusions drawn here are only verified by the database of the compressible channel flows. Whether they hold in wall turbulence with different configurations requires further research. Besides, the wall temperature is deemed to impose non-negligible effects on the energy-containing motions in wall-bounded turbulence. Hence, it is of great significance to investigate them in turbulent boundary layers with different wall temperatures.

\begin{acknowledgments}
We would like to thank Professor Jim\'enez for making the incompressible DNS data available. L.F. acknowledges the fund from the Research Grants Council (RGC) of the Government of Hong Kong Special Administrative Region (HKSAR) with RGC/ECS Project (No. 26200222), the fund from Guangdong Basic and Applied Basic Research Foundation (No. 2022A1515011779), and the fund from the Project of Hetao Shenzhen-Hong Kong Science and Technology Innovation Cooperation Zone (No. HZQB-KCZYB-2020083).

\end{acknowledgments}

\section{APPENDIX A: Scale characteristics at other wall-normal position}\label{OW}
\begin{figure*}
    \centering
    \includegraphics[width=0.45\linewidth]{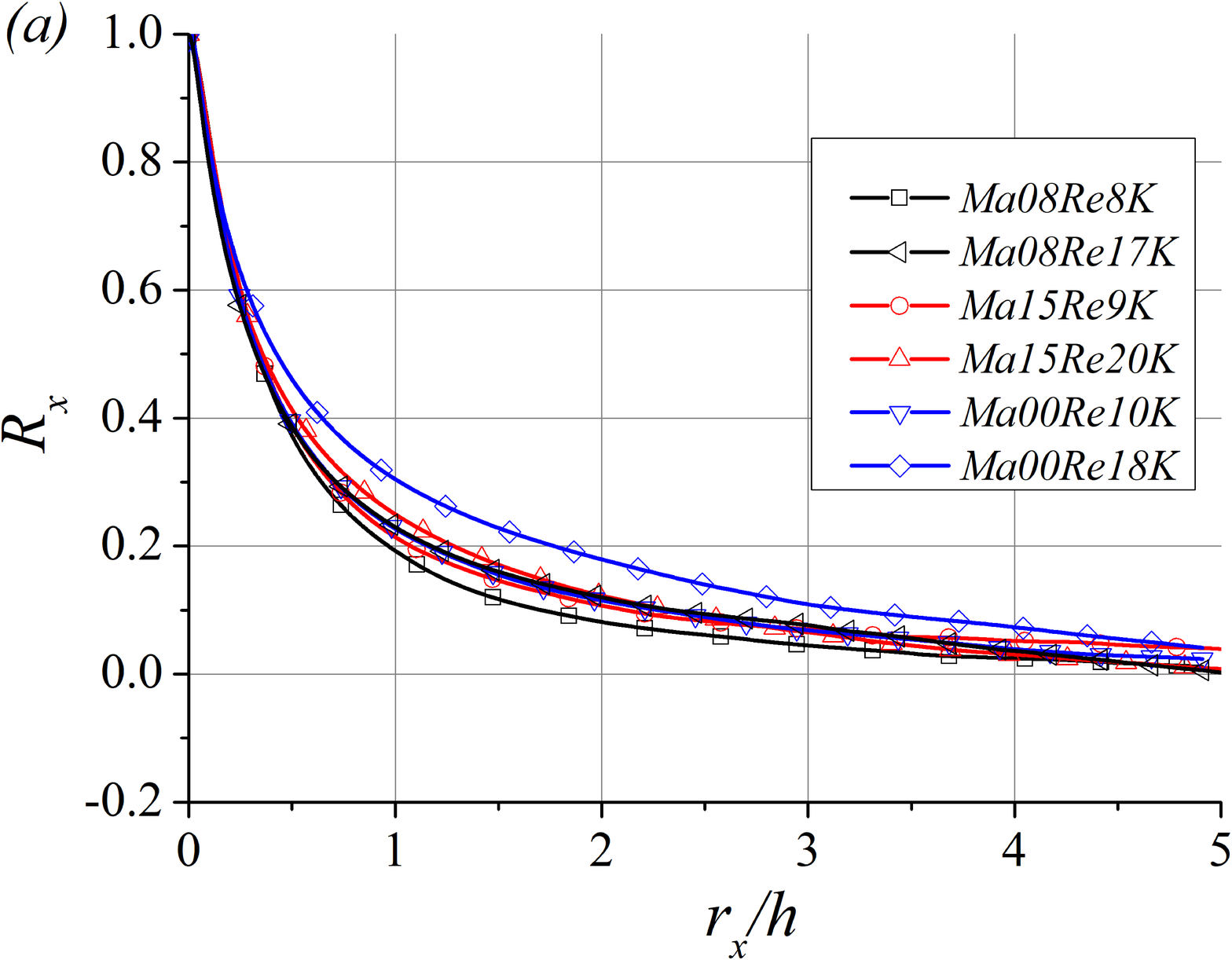}
    \includegraphics[width=0.45\linewidth]{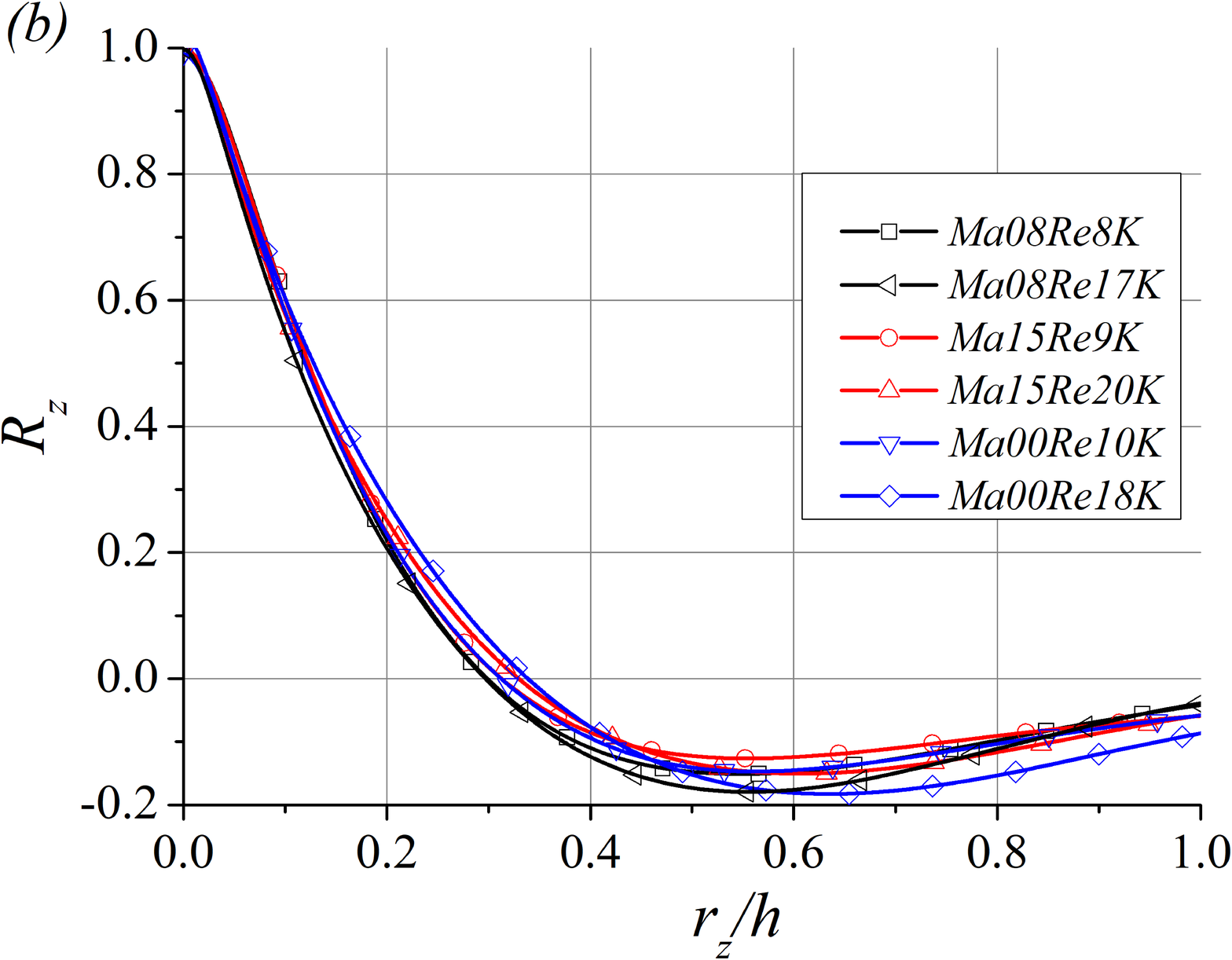}
    \includegraphics[width=0.45\linewidth]{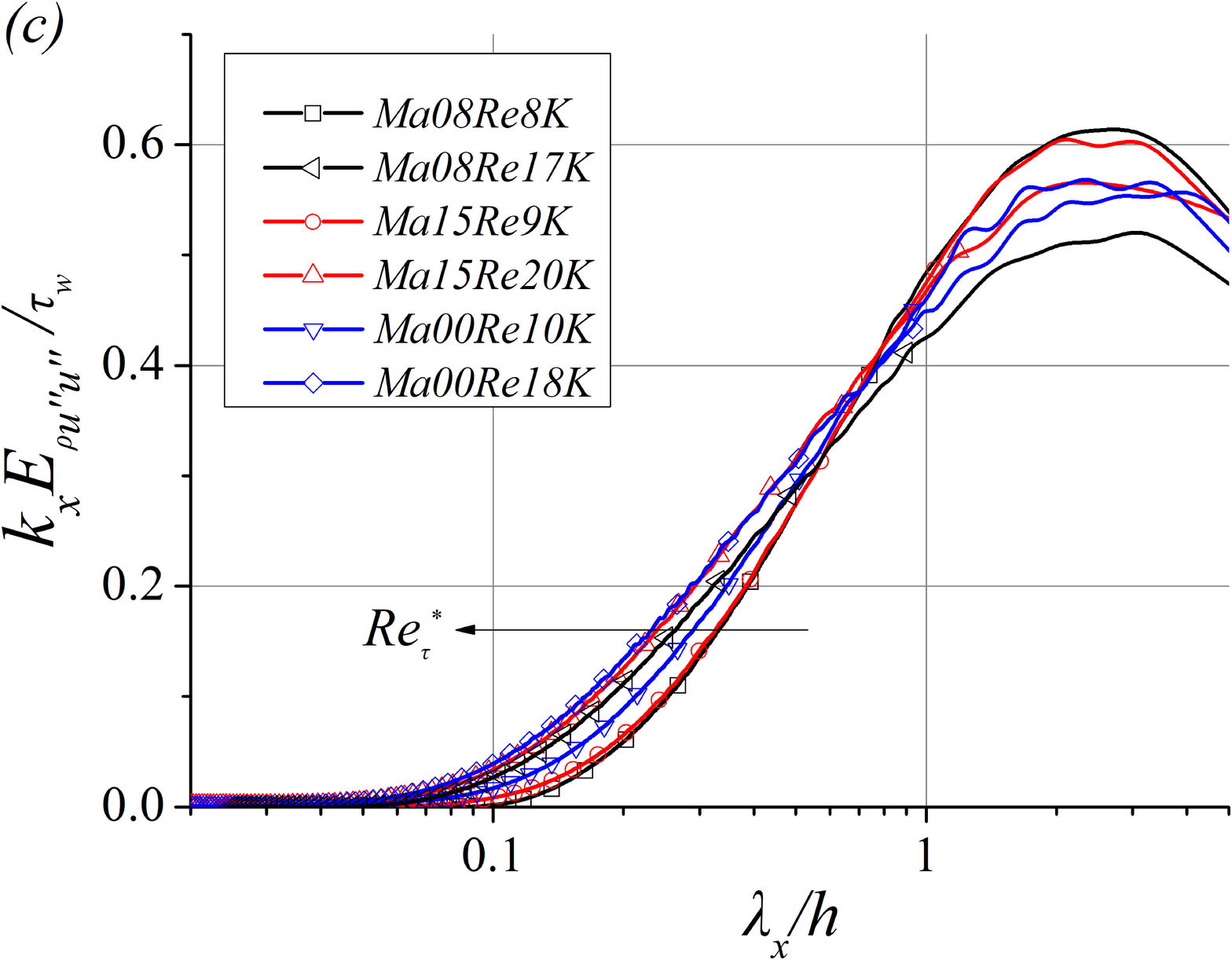}
    \includegraphics[width=0.45\linewidth]{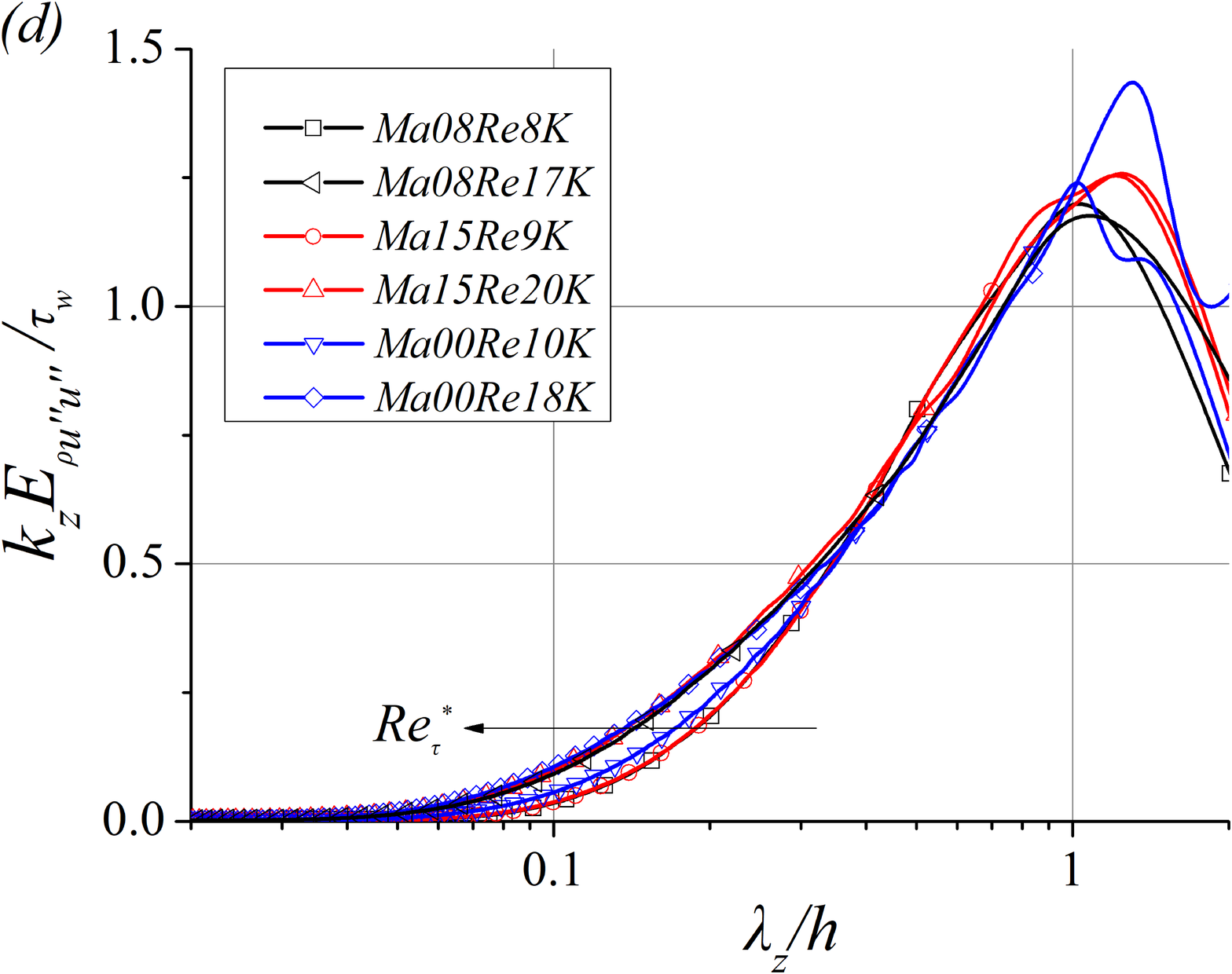}
    \caption{Streamwise ($a$) and  spanwise ($b$) two-point correlations of $\sqrt\rho u^{''}$ for all cases at $y/h=0.3$; streamwise ($c$) and  spanwise ($d$) premultiplied one-dimensional spectra of  $\sqrt\rho u^{''}$ for all cases at $y/h=0.3$.}
    \label{fig:sp5}
\end{figure*}

Fig.~\ref{fig:sp5} shows the  streamwise and  spanwise two-point correlations and spectra of $\sqrt\rho u^{''}$ at $y=0.3h$ for all cases. For the two-point correlations, all the profiles nearly coincide (except the $R_x$ of Ma0018K). For the spectra, the $\lambda_{x,me}$ of
all cases is approximately equal to 2.5, whereas for $\lambda_{z,me}$, that is 1.2. Again, all these suggest that there is no recognizable scale expansion of the outer-region motions in both streamwise and spanwise directions when flow passes the sound barrier.

\bibliography{eddy2}

\end{document}